\pdfoutput=1

\documentclass[aps,prd,showpacs,groupedaddress]{revtex4}               %
\usepackage{hyperref}   
\begin{document}

\preprint{RCTP-9401}

\title{Higher Order Gravity, Gauge Invariant Variables, And
Quantum Behavior In Weyl-Like Geometries}


\author{J.E. Rankin}
\email[]{jrankin@panix.com}
\affiliation{Rankin Consulting, Concord, CA}

\date{July 24, 2018}

\begin{abstract}
This paper presents the detailed, standard treatment of a simple, gauge invariant action for Weyl and Weyl-like Cartan geometries outlined in a previous paper. In addition to the familiar scalar curvature squared and Maxwell terms, the action chosen contains the logarithmic derivative of the scalar curvature combined with the intrinsic four vector (Weyl vector) in a gauge invariant fashion. This introduces higher order derivative terms directly into the action. No separate, ``matter'' fields are introduced. As the usual Weyl metric and four vector are varied, certain gauge invariant combinations of quantities arise naturally as the results are collected, provided the scalar curvature is nonzero. This paper demonstrates the general validity of these results for any gauge choice. Additionally, ``matter'' terms appear in the field equations. Furthermore, the resulting forms isolate the familiar mathematical structure of a coupled Einstein-Maxwell-Schr\"{o}dinger (relativistic) system of classical fields, with the exception of additional, second derivative terms in the stress tensor for the Schr\"{o}dinger field, and the algebraic independence of the conjugate wavefunction. This independence is found to be equivalent to the presence of a second, negative energy, Schr\"{o}dinger field. A detailed comparison is made between this model, and the standard Einstein-Maxwell-Schr\"{o}dinger field theory. The possible use of such continuum models as a basis for quantum phenomena, and some generalizations of the model are discussed.
%

\end{abstract}

\maketitle

\section{Introduction}

In a watershed paper in 1927, London demonstrated that certain intrinsic, quantum mechanical properties can be associated with Weyl geometries with an imaginary fine structure constant\cite{london.orig,adler.london}. More recently, several papers have explored related ideas for both Weyl geometries, and similar geometric structures\cite{rankin.ijtp,galehouse.ijtp.1,%
galehouse.ijtp.2,wheeler.physrev,rankin.caqg,wood.papini}. My own contributions to this list use the concept of gauge invariant variables\cite{rankin.ijtp,rankin.caqg}, a concept which found some use in earlier work as well\cite{adler.london}. In an effort to clarify this concept, this paper presents a particular example of a continuum model handled previously using an action with gauge invariant variables. However, here the action and its variation are initially based on the standard, gauge dependent variables more familiar to workers with Weyl geometries\cite{adler.london}. This derivation follows a procedure outlined in an earlier presentation, but not actually performed there\cite{rankin.caqg}. The purpose is to demonstrate the general validity and utility of results obtained using gauge invariant variables, provided the scalar curvature of the structure is nonzero. Indeed, use of such variables often reduces the labor of a given derivation, and provides simple paths to construct, interpret, and generalize models when the standard, gauge dependent variables would not do so.

In this case, it immediately isolates very nearly exactly the mathematical structure of coupled, relativistic Einstein-Maxwell-Schr\"{o}dinger fields, even though {\it no} Schr\"{o}dinger (Klein-Gordon) fields are independently introduced or varied (``Schr\"{o}dinger'' will mean ``relativistic Schr\"{o}dinger'', or ``Klein-Gordon'', throughout this paper, and will be freely interchanged with the term ``Klein-Gordon''). This result is seen to be related to a kinematic tendency of such geometries. However, choosing this action, which is closest in form to the original Einstein-Maxwell form of action when gauge invariant variables are used, greatly enhances this tendency. The results then reproduce the full structure of the coupled 
Einstein-Maxwell-Schr\"{o}dinger field equations, except for additional terms in the stress tensor for the Schr\"{o}dinger fields, and the absence of an algebraic constraint between the wavefunction and the conjugate wavefunction which obeys the charge conjugate wave equation. This last aspect will be found to be equivalent to the appearance of a second, negative energy Schr\"{o}dinger field. In order to fully produce the complex valued character of the Schr\"{o}dinger fields, the electromagnetic field is made purely imaginary, reintroducing London's original, imaginary fine structure constant. Nevertheless, what is essentially a simple unification of standard, Einstein-Maxwell classical field theory, in a Weyl or Weyl-like, complex valued geometry, basically also automatically produces the forms of the Schr\"{o}dinger field, current, and stress-tensor, all on the level of a classical continuum theory. The choice of an action capable of producing the Einstein-Maxwell free fields exactly, rather than the slightly different results of Weyl, is quite important for this result.

Once these points are made, this paper refocuses on the possible significance of the intrinsic Schr\"{o}dinger-like behavior in classical geometries found in this and similar cases, and its potential as a framework for quantum phenomena. A detailed comparison is made between the classical continuum model which appears here, and the standard theory of the coupled 
Einstein-Maxwell-Schr\"{o}dinger field, with its related quantum mechanics of particles. The differences are discussed in order of ascending importance or difficulty, beginning with the additional terms in the stress tensor, which are seen to be important only in strong gravitational fields, or for the overall magnitude of the gravitational field. Next, the question of positivity of energy is examined, and it will be seen that the existence of some antigravitating solutions currently remains an open possibility. However, any such solutions which are internally self consistent will be seen at least to possess no simple symmetry with gravitating solutions. Therefore they are not required to have similar properties or chances of existence, unlike the case with antimatter versus matter solutions. Finally, the question of particlelike phenomena is raised. It will be seen that the model {\it cannot} assume the a priori existence of discrete (singular) particles {\it in addition} to the fields without losing the bulk of the Schr\"{o}dinger behavior of interest. Furthermore, second quantization of its fields lacks harmony with this entire approach, and may be difficult in practice, although it would be the conventional answer to this question. Thus, production of the particle aspect of observed, wave-particle duality becomes the primary challenge for the model. In this regard, the ``minor'', additional terms in the stress tensor, and general relativistic modifications to the Schr\"{o}dinger field equations are seen 
%
%
to combine to produce simple, nonlinear Schr\"{o}dinger equation forms in strong fields. Other nonlinear effects are also present then, but in microscopic physics, all of these effects would normally be expected to become important only on scales several orders of magnitude smaller than $10^{-17}$ centimeters.

The paper concludes with a brief survey of generalizations aimed at including spin, variability of the ``rest mass'' parameter in the Schr\"{o}dinger equations, and additional interactions, possible including some related to SU(2). These points are all related in this approach.

\section{A Simple Case Study}
\label{case.study}

\subsection{Preliminaries}

Take the action presented in section 3 of an earlier work\cite{rankin.caqg}. This was presented there as a functional of real, gauge invariant variables in either a pure Weyl geometry, or a closely related Weyl-like Cartan geometry, both with nonzero scalar curvature. The Weyl-like Cartan geometry is developed in section 1 of the same reference, and either geometry leads to the same action and equations of motion. However, the Weyl-like Cartan geometry replaces the nonzero nonmetricity of the pure Weyl case with a nonzero torsion with no tracefree part\cite{rankin.caqg,wheeler.rankin.poster}. In either case, the action presented was written essentially as
\begin{equation}
I=\int {[(\hat R-2\sigma )-{\textstyle{1 \over 2}}\, j^2(\hat p_{\mu \nu }
\hat p^{\mu \nu })+\hat \beta (\hat R+6\hat v^\mu _{\; \; ||\mu }
+6\hat v^\mu \hat v_\mu -1)]\sqrt {-\hat g}\, d^4x}
\label{orig.gi.action}
\end{equation}
Here, the more conventional use of Greek subscripts and superscripts is adopted rather than the earlier use of Latin ones. The rest of the notation is explained in the earlier reference, but will be reintroduced into this 
paper later as needed. The above is provided here only to pinpoint the example being detailed.

Now it was noted in the same section in the earlier paper that the action could be reexpressed in terms of the conventional, gauge varying metric $g_{\mu \nu }$ (Weyl metric), and four vector $v_{\mu }$ (Weyl vector)\cite{rankin.caqg}. Let 
$\{ ^{\: \mu }_{\nu \alpha } \}$ be the Christoffel symbol constructed from 
$g_{\mu \nu }$, and let $\Gamma ^\mu _{\nu \alpha }$ be the full Weyl connection
\begin{equation}
\Gamma ^\mu _{\nu \alpha }=\{ ^{\: \mu }_{\nu \alpha } \}+\delta ^\mu _\nu 
v_\alpha +\delta ^\mu _\alpha v_\nu -g_{\nu \alpha }v^\mu 
\label{def.connect.weyl}
\end{equation}
or the Weyl-like Cartan connection
\begin{equation}
\Gamma ^\mu _{\nu \alpha }=\{ ^{\: \mu }_{\nu \alpha } \}+
\delta ^\mu _\alpha v_\nu -g_{\nu \alpha }v^\mu 
\label{def.connect.cartan}
\end{equation}
Then use $,\mu$ to denote the partial derivative with respect to coordinate 
$x^\mu $, and define the full curvature tensor,
\begin{equation}
B_{\nu \alpha \gamma }^\mu =\Gamma _{\nu \gamma ,\alpha }^\mu -
\Gamma _{\nu \alpha ,\gamma }^\mu +\Gamma _{\nu \gamma }^\lambda 
\Gamma _{\lambda \alpha }^\mu -\Gamma _{\nu \alpha }^\lambda 
\Gamma _{\lambda \gamma }^\mu 
\label{def.full.curv}
\end{equation}
Additionally, assume that all coordinates $x^\mu $ are dimensionless, dividing them by a basic length or scale factor if necessary to render them so. Denote that scale factor by $b^{-1/2}_0$. The importance of this scale factor will become clearer later.

Define the analogous quantity using $\{ ^{\: \mu }_{\nu \alpha } \}$ in 
place of $\Gamma ^\mu _{\nu \alpha }$ as $R_{\nu \alpha \gamma }^\mu $, the Riemannian portion of the curvature. The contracted tensors are
$B_{\nu \alpha } =B_{\nu \alpha \mu }^\mu $, and 
$R_{\nu \alpha } =R_{\nu \alpha \mu }^\mu $, and the respective scalar curvatures are $B =g^{\nu \alpha } B_{\nu \alpha }$, and 
$R =g^{\nu \alpha } R_{\nu \alpha }$.

Now it is characteristic of these geometries that $B_{\nu \alpha 
\gamma }^\mu $ is invariant under the gauge transformation
\begin{equation}
\bar g_{\mu \nu} =sg_{\mu \nu }
\label{gauge.law.g}
\end{equation}
and
\begin{equation}
\bar v_\mu =v_\mu - ({\textstyle{1 \over 2}}\, ln\, s)_{,\mu }
\label{gauge.law.v}
\end{equation}
where $s$ is a nonzero scalar function
\cite{adler.london,rankin.ijtp,rankin.caqg}. Therefore the contracted tensor $B_{\nu \alpha }$ is also gauge invariant, and the scalar $B$ transforms via $\bar B =(B/s)$ under this transformation, since
\begin{equation}
\bar g^{\mu \nu} =(1/s)\, g^{\mu \nu }
\label{gauge.law.invrs.g}
\end{equation}
Thus, the quantity $B^2 \sqrt {-g}$ is gauge invariant, as Weyl noted in some of the original work on such geometries\cite{weyl.stm}. This is a standard starting point for the gravitational part of the action in these geometries\cite{weyl.stm,eddington.mtr,adler.london}.

The second standard term introduced into the action is the electromagnetic term\cite{weyl.stm,eddington.mtr,adler.london,rankin.caqg}. It will be found advantageous shortly to let the Weyl vector $v_\mu $ be proportional to the usual electromagnetic potential $A_\mu $, rather than exactly equal to it, so denote the curl of the Weyl vector here by 
$p_{\mu \nu }$ rather than $F_{\mu \nu }$ to aid in tracking units {\it(after all, units are what gauge theory is originally about)}, so
\begin{equation}
p_{\mu \nu }=v_{\nu ,\mu }-v_{\mu ,\nu }
\label{def.unhatted.p}
\end{equation}
The combination of the proportionality constant and use of dimensionless coordinates will be chosen to render $p_{\mu \nu }$ itself dimensionless, so that the standard electromagnetic term in the action finally becomes 
$-{\textstyle{1 \over 2}}\, j^2 p_{\mu \nu }p^{\mu \nu } \sqrt {-g}\, $. The constant $j$ is itself ultimately dimensionless, and its square is used to allow it to be directly absorbed into $p_{\mu \nu }$ if that should prove desirable.

But now, a third term will be added to the usual action. This term, which will be justified by the results produced, is 
$-(6/C)\, Bg^{\mu \nu }\, 
[v_\mu - ({\textstyle{1 \over 2}}\, ln\, B)_{,\mu }]
[v_\nu - ({\textstyle{1 \over 2}}\, ln\, B)_{,\nu }] \sqrt {-g}\, $, where the $C$ is another (nonzero) dimensionless constant. A careful check will reveal that this term is also gauge invariant, like the preceding two terms. However, it {\it does} contain a logarithmic derivative of $B$, and so introduces higher order derivatives into the action.

To produce the final action to be examined, introduce yet another constant into the first term, giving
{\samepage 
\begin{eqnarray}
I & = & \int {\{ [(C-2\sigma )/C^2]\, B^2 
-{\textstyle{1 \over 2}}\, j^2 p_{\mu \nu }p^{\mu \nu }} \nonumber \\
  &   & -(6/C)\, Bg^{\mu \nu }\, 
[v_\mu - ({\textstyle{1 \over 2}}\, ln\, B)_{,\mu }]
[v_\nu - ({\textstyle{1 \over 2}}\, ln\, B)_{,\nu }]\} \sqrt {-g}\, d^4 x
\label{example.main.action}
\end{eqnarray}}%
Here, the C in the first and last terms is the same constant, $\sigma $ is another dimensionless constant, and it is assumed that $B \neq 0$. Except for the constant $C$, this is the action outlined in a previous paper as being equivalent to the action of equation (\ref{orig.gi.action})\cite{rankin.caqg}. Proving this is one object of this paper.

To make bookkeeping easier in the derivation of results from the action of equation (\ref{example.main.action}), now {\it define}
{\samepage 
\begin{eqnarray}
\hat v_\mu & = & v_\mu - [{\textstyle{1 \over 2}}\, ln\, (B/C)]_{,\mu }
\nonumber \\
           & = & v_\mu - ({\textstyle{1 \over 2}}\, ln\, B)_{,\mu }
\label{def.vhat}
\end{eqnarray}}%
Besides simplifying notation, equation (\ref{def.vhat}) defines a {\it gauge invariant} quantity. As long as $B\neq 0$, which is a gauge invariant inequality, it is defined in {\it any} gauge, and has the same value regardless of the gauge. This has been noted in previous work\cite{adler.london,rankin.ijtp,rankin.caqg}.

Now equation (\ref{def.vhat}) has a simple analog for the metric tensor. Define
\begin{equation}
\hat g_{\mu \nu }=(B/C)\, g_{\mu \nu }
\label{def.ghat}
\end{equation}
Like $\hat v_\mu $, $\hat g_{\mu \nu }$ is defined for any gauge, and has the same value regardless of the gauge. Using it, it is possible to define a gauge invariant $\hat g^{\mu \nu }$ such that 
$\hat g^{\mu \nu }\hat g_{\nu \lambda }=\delta ^\mu _\lambda $, and a gauge invariant Christoffel symbol 
$\{ ^{\: \hat \mu }_{\nu \alpha } \}$, each defined by
\begin{equation}
\hat g^{\mu \nu }=(C/B)\, g^{\mu \nu }
\label{def.ghat.inv}
\end{equation}
and
\begin{equation}
\{ ^{\: \hat \mu }_{\nu \alpha } \} = {\textstyle{1 \over 2}}\, 
\hat g^{\mu \lambda }[\hat g_{\nu \lambda ,\alpha }+
\hat g_{\alpha \lambda ,\nu }-\hat g_{\nu \alpha ,\lambda }]
\label{def.gi.chrst.symbol}
\end{equation}
This in turn allows definition of a gauge invariant Riemannian portion to the overall curvature tensor via
\begin{equation}
\hat R^\mu _{\nu \alpha \gamma }=
\{ ^{\: \hat \mu }_{\nu \gamma } \} _{,\alpha}-
\{ ^{\: \hat \mu }_{\nu \alpha } \} _{,\gamma}+
\{ ^{\: \hat \lambda }_{\nu \gamma } \} 
\{ ^{\: \hat \mu }_{\lambda \alpha } \}-
\{ ^{\: \hat \lambda }_{\nu \alpha } \} 
\{ ^{\: \hat \mu }_{\lambda \gamma } \}
\label{def.gi.r.curv}
\end{equation}
Then define $\hat R_{\nu \alpha }=\hat R^\mu _{\nu \alpha \mu }$, and 
$\hat R = \hat g^{\nu \alpha }\hat R_{\nu \alpha }$. Also define 
$\hat p_{\mu \nu }=p_{\mu \nu }$, since $p_{\mu \nu }$ is already gauge invariant, and note that the curl of $\hat v_\mu $ equals the curl of 
$v_\mu $. With these, one need only add that the $||$ derivative is the covariant derivative with respect to the gauge invariant Christoffel symbol of equation (\ref{def.gi.chrst.symbol}), and that all hatted quantities have their indices raised and lowered using $\hat g^{\mu \nu}$ and 
$\hat g_{\mu \nu}$, and the notation used in equation (\ref{orig.gi.action}) is now in place ($\hat \beta $ is a gauge invariant Lagrange multiplier). The notation has also considerably outpaced its relevance in this paper, as well as explanations about its use, so it is time to return to equation (\ref{example.main.action}).

\subsection{Equations of Motion}

The action of equation (\ref{example.main.action}) is considered to be a functional of just two quantities, the Weyl metric $g^{\mu \nu }$, and the Weyl vector $v_\mu $ (both are still referred to as Weyl variables here because of their gauge properties, even in the Cartan case). Note that no independent matter or wavefunction terms are present in this purely geometric, field action. At this point, it appears simply as another action for a geometrically unified, classical field theory solely of gravitation and electromagnetism, similar to Weyl's earlier attempt\cite{weyl.stm}.

The variation of the action will be done in several stages, the first of which leads simply to
{\samepage 
\begin{eqnarray}
\delta I & = & \int {\{ [j^2(p_\mu ^{\; \; \gamma }p_{\gamma \nu }+
{\textstyle{1 \over 4}}\, g_{\mu \nu }p_{\gamma \sigma }
p^{\gamma \sigma })-{\textstyle{1 \over 2}}(C-2\sigma )
(B^2 /C^2 )\, g_{\mu \nu }-(6/C)B\hat v_\mu \hat v_\nu }
\nonumber \\
         &   & +(3/C)Bg^{\gamma \sigma }\hat v_\gamma 
\hat v_\sigma g_{\mu \nu }]\sqrt {-g}\, \delta g^{\mu \nu }
-[2j^2 (\sqrt {-g}\, p^{\mu \nu })_{,\nu }+12(B/C)\, g^{\mu \nu }
\hat v_\nu \sqrt {-g}\, ]\delta v_\mu \nonumber \\
         &   & +[2(C-2\sigma )(B/C^2)\sqrt {-g}-(6/C)\, 
g^{\gamma \sigma }\hat v_\gamma \hat v_\sigma \sqrt {-g}
\nonumber \\
         &   & -(6/C)(1/B)(B\sqrt {-g}\, g^{\gamma \sigma }
\hat v_\gamma )_{,\sigma }]\delta B\} d^4 x
\label{var.1}
\end{eqnarray}}%
Here, the quantity $\hat v_\mu $ has already been introduced, but so far only for convenience to compress the notation {\it after} the variations are performed. No gauge specialization has been invoked. Of course, $B$ is itself a function of $g^{\mu \nu }$ and $v_\mu $, so the variation is still incomplete because the $\delta B$ term must be evaluated yet.

To evaluate the $\delta B$ term, note that in both the Weyl and the 
Weyl-like Cartan geometry, it is true that in all gauges that
\begin{equation}
B=R+6v^\mu _{\; \; ;\mu }+6v^\mu v_\mu
\label{basic.identity}
\end{equation}
where the $;$ derivative denotes the usual covariant derivative with respect to the ordinary Christoffel symbols. This is well known for the Weyl case\cite{weyl.stm}, and can be easily established for the Weyl-like Cartan case by direct substitution of equation (\ref{def.connect.cartan}) into equation (\ref{def.full.curv}), followed by the appropriate contractions\cite{rankin.caqg}.

Since this shows that $\delta B$ will involve terms like 
$g^{\mu \nu }\delta R_{\mu \nu }$ in the intermediate steps, and since these and other terms will eventually generate derivatives of the current coefficient of $\delta B$, that coefficient should be simplified as much as possible before proceeding. However, to keep the result fully general, 
{\it no} gauge fixing should be invoked.

To do this, first notice that in {\it any} gauge, $R$ and $\hat R$ are related by the equation
\begin{equation}
R=(B/C)\, \hat R-(6/\sqrt {-g}\, )[\sqrt {-g}\, g^{\mu \nu }
({\textstyle{1 \over 2}}\, ln\, B)_{,\nu }]_{,\mu }-6g^{\mu \nu }
({\textstyle{1 \over 2}}\, ln\, B)_{,\mu }
({\textstyle{1 \over 2}}\, ln\, B)_{,\nu }
\label{rhat.to.r}
\end{equation}
This relation may be derived by first substituting for $\hat g_{\mu \nu }$ and $\hat g^{\mu \nu }$ in equation (\ref{def.gi.chrst.symbol}) using equations (\ref{def.ghat}) and (\ref{def.ghat.inv}). This gives
\begin{equation}
\{ ^{\: \hat \mu }_{\nu \alpha } \} = \{ ^{\: \mu }_{\nu \alpha } \}
+\delta ^\mu _\nu ({\textstyle{1 \over 2}}\, ln\, B)_{,\alpha }
+\delta ^\mu _\alpha ({\textstyle{1 \over 2}}\, ln\, B)_{,\nu }
-g_{\nu \alpha }g^{\mu \gamma }
({\textstyle{1 \over 2}}\, ln\, B)_{,\gamma }
\label{chrsthat.to.chrst}
\end{equation}
This may be substituted into equation (\ref{def.gi.r.curv}), and appropriate contractions performed to get equation (\ref{rhat.to.r}). Also, one of the intermediate steps in this process will give
{\samepage 
\begin{eqnarray}
\hat R_{\nu \alpha } & = & R_{\nu \alpha }+
2({\textstyle{1 \over 2}}\, ln\, B)_{;\nu ;\alpha}+
g_{\nu \alpha }g^{\gamma \sigma }
({\textstyle{1 \over 2}}\, ln\, B)_{;\gamma ;\sigma}\nonumber \\
                     &   & -2({\textstyle{1 \over 2}}\, ln\, B)_{,\nu }
({\textstyle{1 \over 2}}\, ln\, B)_{,\alpha }+
2g_{\nu \alpha }g^{\gamma \sigma }
({\textstyle{1 \over 2}}\, ln\, B)_{,\gamma }
({\textstyle{1 \over 2}}\, ln\, B)_{,\sigma }
\label{rhatnualpha.to.rnualpha}
\end{eqnarray}}%
This latter equation will be needed later.

Now substitute for $R$ in equation (\ref{basic.identity}) using equation (\ref{rhat.to.r}). Using the definitions of equations (\ref{def.vhat}), (\ref{def.ghat}), and (\ref{def.ghat.inv}), and collecting terms, one obtains
\begin{equation}
B=(B/C)[\hat R+(6/\sqrt {-\hat g}\,)(\sqrt {-\hat g}\, \hat g^{\mu \nu }
\hat v_\nu )_{,\mu }+6\hat g^{\mu \nu }\hat v_\mu \hat v_\nu ]
\label{basic.identity.2}
\end{equation}
Since $B \neq 0$, this immediately gives
\begin{equation}
C=\hat R+(6/\sqrt {-\hat g}\,)(\sqrt {-\hat g}\, \hat g^{\mu \nu }
\hat v_\nu )_{,\mu }+6\hat g^{\mu \nu }\hat v_\mu \hat v_\nu 
\label{gi.basic.identity}
\end{equation}
This is nothing more or less than another form of equation (\ref{basic.identity}). Indeed, direct substitution into this equation from equations (\ref{def.vhat}), (\ref{def.ghat}), and (\ref{def.ghat.inv}), will simply reproduce the original equation (\ref{basic.identity}), so that the entire derivation of equation (\ref{gi.basic.identity}) is completely reversible. {\it No terms have been lost, and no gauge has been fixed.} This is an important point, since equation (\ref{gi.basic.identity}) does resemble a gauge condition, but it is not one. Among other things, this equation is useful for converting from an action in gauge invariant variables, such as equation (\ref{orig.gi.action}), into an action in the Weyl variables, such as equation (\ref{example.main.action}). This is done by substituting for $\hat R$ from equation (\ref{gi.basic.identity}) in the action in gauge invariant variables, then expanding all the remaining terms using the definitions of the hatted variables. Of course, the particular example of equation (\ref{orig.gi.action}) assumed $C=1$.

Now simplify equation (\ref{var.1}) by likewise merging terms into their gauge invariant forms in the coefficients of the separate variations. This gives
{\samepage 
\begin{eqnarray}
\delta I & = & \int {\{ [j^2(\hat p_\mu ^{\; \; \gamma }
\hat p_{\gamma \nu }+{\textstyle{1 \over 4}}\, \hat g_{\mu \nu }
\hat p_{\gamma \sigma }\hat p^{\gamma \sigma })-
{\textstyle{1 \over 2}}(C-2\sigma )\, \hat g_{\mu \nu }-6\hat v_\mu 
\hat v_\nu }\nonumber \\
         &   & +3\hat v^\gamma \hat v_\gamma \hat g_{\mu \nu }]
(C/B)\sqrt {-\hat g}\, \delta g^{\mu \nu }
-[2j^2 (\sqrt {-\hat g}\, \hat p^{\mu \nu })_{,\nu }+
12\hat v^\mu \sqrt {-\hat g}\, ]\delta v_\mu \nonumber \\
         &   & +[2(C-2\sigma )\sqrt {-\hat g}-6\hat v^\gamma 
\hat v_\gamma \sqrt {-\hat g}-6(\sqrt {-\hat g}\, 
\hat v^\sigma )_{,\sigma }](1/B)\delta B\} d^4 x
\label{var.2}
\end{eqnarray}}%
where indices on hatted quantities are raised and lowered using the hatted metric, not the gauge dependent metric. Now apply equation (\ref{gi.basic.identity}) to the coefficient of $\delta B$ to get
{\samepage 
\begin{eqnarray}
\delta I & = & \int {\{ [j^2(\hat p_\mu ^{\; \; \gamma }
\hat p_{\gamma \nu }+{\textstyle{1 \over 4}}\, \hat g_{\mu \nu }
\hat p_{\gamma \sigma }\hat p^{\gamma \sigma })-
{\textstyle{1 \over 2}}(C-2\sigma )\, \hat g_{\mu \nu }-6\hat v_\mu 
\hat v_\nu }\nonumber \\
         &   & +3\hat v^\gamma \hat v_\gamma \hat g_{\mu \nu }]
(C/B)\sqrt {-\hat g}\, \delta g^{\mu \nu }
-[2j^2 (\sqrt {-\hat g}\, \hat p^{\mu \nu })_{,\nu }+
12\hat v^\mu \sqrt {-\hat g}\, ]\delta v_\mu \nonumber \\
         &   & +[(\hat R-4\sigma +C)\sqrt {-\hat g}\, ](1/B)\delta B\} d^4 x
\label{var.3}
\end{eqnarray}}%
Of course, the leftover $(1/B)$ term outside the brackets in the 
$\delta B$ term will still complicate the remaining reduction of this term into the correct $\delta g^{\mu \nu }$ and $\delta v_\mu $ terms.

To proceed, define
\begin{equation}
\hat \beta =(1/C)(\hat R-4\sigma )
\label{def.betahat}
\end{equation}
and use equation (\ref{basic.identity}) to write
{\samepage 
\begin{eqnarray}
\delta B & = & R_{\mu \nu }\delta g^{\mu \nu }+
g^{\mu \nu }\delta R_{\mu \nu }+(3/\sqrt {-g}\, )(\sqrt {-g}\, 
g^{\alpha \gamma }v_\gamma )_{,\alpha }g_{\mu \nu }\delta 
g^{\mu \nu }\nonumber \\
         &   & +(6/\sqrt {-g}\, )[\delta (\sqrt {-g}\, g^{\gamma \sigma }
v_\sigma )]_{,\gamma }+6v_\mu v_\nu \delta g^{\mu \nu }+
12g^{\mu \nu }v_\nu \delta v_\mu 
\label{delta.b.exp}
\end{eqnarray}}%
Note that all of the terms in equation (\ref{delta.b.exp}) involve only the gauge dependent, Weyl variables. Of course, terms like 
$\delta R_{\mu \nu }$ require still further reduction, but by well known procedures\cite{adler.london}.

After a long, tedious, but straightforward process, including multiple use of all of the previously developed relations between hatted and unhatted quantities, even equation (\ref{rhatnualpha.to.rnualpha}), the result can be written
{\samepage 
\begin{eqnarray}
\delta I & = & \int {\{ [j^2(\hat p_\mu ^{\; \; \gamma }
\hat p_{\gamma \nu }+{\textstyle{1 \over 4}}\, \hat g_{\mu \nu }
\hat p_{\gamma \tau }\hat p^{\gamma \tau })-{\textstyle{1 \over 2}}
(C-2\sigma )\, \hat g_{\mu \nu }}+3\hat v^\gamma \hat v_\gamma 
\hat g_{\mu \nu }\nonumber \\
         &   & +\hat \beta _{||\mu ||\nu }-3(\hat \beta _{,\mu }\hat v_\nu +
\hat \beta _{,\nu }\hat v_\mu )+3\hat \beta _{,\gamma }\hat v^\gamma 
\hat g_{\mu \nu }+6\hat \beta \hat v_\mu \hat v_\nu 
-\hat \beta _{||\gamma ||\tau }\hat g^{\gamma \tau }
\hat g_{\mu \nu }\nonumber \\
         &   & +3\hat \beta \hat v^\gamma _{\; \; ||\gamma }
\hat g_{\mu \nu }+3\hat v^\gamma _{\; \; ||\gamma }
\hat g_{\mu \nu }+\hat \beta \hat R_{\mu \nu }+\hat R_{\mu \nu }]
(C/B)\sqrt {-\hat g}\, \delta g^{\mu \nu }\nonumber \\
         &   & -[2j^2 \hat p^{\mu \nu }_{\; \; \; \; ||\nu }-
12\hat \beta \hat v^\mu +6\hat g^{\mu \nu }\hat \beta _{,\nu }]
\sqrt {-\hat g}\, \delta v_\mu \} d^4 x
\label{var.result}
\end{eqnarray}}%
The $(C/B)$ term may be ignored in the $\delta g^{\mu \nu }$ term, since 
$B \neq 0$ by assumption. Then applying equation (\ref{gi.basic.identity}) twice more in the $\delta g^{\mu \nu }$ term, the results are finally
{\samepage 
\begin{eqnarray}
\hat R_{\mu \nu }-{\textstyle{1 \over 2}}\, \hat R\hat g_{\mu \nu }+
\sigma \hat g_{\mu \nu } & = & -j^2(\hat p_\mu ^{\; \; \gamma }
\hat p_{\gamma \nu }+{\textstyle{1 \over 4}}\, \hat g_{\mu \nu }
\hat p_{\gamma \tau }\hat p^{\gamma \tau })-\hat \beta _{||\mu ||\nu }
\nonumber \\
 &   & +3(\hat \beta _{,\mu }\hat v_\nu +
\hat \beta _{,\nu }\hat v_\mu )-3\hat \beta _{,\gamma }\hat v^\gamma 
\hat g_{\mu \nu }-6\hat \beta \hat v_\mu \hat v_\nu \nonumber \\
 &   & +\hat \beta _{||\gamma ||\tau }\hat g^{\gamma \tau }
\hat g_{\mu \nu }+3\hat \beta \hat v^\gamma \hat v_\gamma 
\hat g_{\mu \nu }-{\textstyle{1 \over 2}}\, C\hat \beta 
\hat g_{\mu \nu }\nonumber \\
 &   & -\hat \beta (\hat R_{\mu \nu }-{\textstyle{1 \over 2}}\, 
\hat R\hat g_{\mu \nu })
\label{gi.einstein.result}
\end{eqnarray}}%
and
\begin{equation}
2j^2 \hat p^{\mu \nu }_{\; \; \; \; ||\nu }=
-6\hat g^{\mu \nu }\hat \beta _{,\nu }+12\hat \beta \hat v^\mu 
\label{gi.maxwell.result}
\end{equation}
Except for the factor $C$, these are exactly the same results obtained in section 3 of the earlier paper from equation (\ref{orig.gi.action}) considered as a functional of $\hat g^{\mu \nu }$, $\hat v_\mu $, and 
$\hat \beta $\cite{rankin.caqg}. In fact in the earlier work, $C=1$ was used, so if the action of equation (\ref{orig.gi.action}) is replaced by
\begin{equation}
I=\int {[(\hat R-2\sigma )-{\textstyle{1 \over 2}}\, j^2(\hat p_{\mu \nu }
\hat p^{\mu \nu })+\hat \beta (\hat R+6\hat v^\mu _{\; \; ||\mu }
+6\hat v^\mu \hat v_\mu -C)]\sqrt {-\hat g}\, d^4x}
\label{cmod.gi.action}
\end{equation}
this will bring the results into exact agreement. However, here the results have been obtained instead from the independent variation of the original Weyl variables in equation (\ref{example.main.action}). At {\it no} point in this derivation has any particular gauge been fixed. Equations (\ref{gi.einstein.result}) and (\ref{gi.maxwell.result}) are still entirely general as regards choice of gauge, even if they are somewhat compressed in their expression via the use of gauge invariant variables. Just as equation (\ref{gi.basic.identity}) can be expanded into equation (\ref{basic.identity}) by substitution from equations (\ref{def.vhat}), (\ref{def.ghat}), and (\ref{def.ghat.inv}), so can these equations be fully expanded into expressions in the gauge dependent, Weyl variables by the same substitutions, as well as use of equation (\ref{def.betahat}).

\subsection{Advantages of Gauge Invariant Variables}

Since the form of equation (\ref{gi.einstein.result}) is suggestive as it stands, why not consider equations (\ref{gi.einstein.result}) and (\ref{gi.maxwell.result}) in their current form? Certainly, use of these gauge invariant variables in action principle (\ref{cmod.gi.action}) will yield the above equations of motion with a fraction of the effort expended above\cite{rankin.caqg}. And, action principle (\ref{cmod.gi.action}) has a simpler form, since it appears as just a standard Einstein-Maxwell action\cite{adler.london}, but with the intrinsic constraint of equation (\ref{gi.basic.identity}) added using a Lagrange multiplier because the gauge invariant variables are not independent\cite{rankin.caqg}. But since equation (\ref{basic.identity}) is an intrinsic feature of these geometries, it and its gauge invariant form, equation (\ref{gi.basic.identity}), are also part of the results, even when the longer derivation of this paper is used. Nor is equation (\ref{def.betahat}) lost anywhere, since it emerges from the trace of equation (\ref{gi.einstein.result}), conservation of charge in equation (\ref{gi.maxwell.result}), and equation (\ref{gi.basic.identity}). The two derivations are completely equivalent in resulting equations, which is as it should be. After all, equation (\ref{example.main.action}) really is equation (\ref{cmod.gi.action}) rephrased into the standard, Weyl variables\cite{rankin.caqg}.

One immediate consequence of this approach is less potential ambiguity with boundary conditions. Since the variables so far are all gauge invariant, particular values on boundaries are not disturbed by the gauge freedom. For the same reason, the gauge invariant variables may be more easily related to actual physical observables, since such observables should also be independent of choice of gauge\cite{wheeler.scalars}.

But the most interesting justification for use of gauge invariant variables probably comes from actual insight into the mathematical structure of these forms as it relates to existing, familiar physical theories. Consider equation (\ref{gi.basic.identity}), and expand {\it just the Weyl vector part} via equation (\ref{def.vhat}). This amounts to a return to the familiar, gauge dependent electromagnetic potential, with full freedom to arbitrarily set its gauge. The result is
{\samepage 
\begin{eqnarray}
& (1/\sqrt {-\hat g}\, ) & \{ \sqrt {-\hat g}\, \hat g^{\mu \nu }[v_\nu -
({\textstyle{1 \over 2}}\, ln\, B)_{,\nu }]\} _\mu \nonumber \\
&  & +\hat g^{\mu \nu }[v_\mu -({\textstyle{1 \over 2}}\, 
ln\, B)_{,\mu }][v_\nu -({\textstyle{1 \over 2}}\, ln\, B)_{,\nu }]=
{\textstyle{1 \over 6}}\, (C-\hat R)
\label{intermed.wave.eq}
\end{eqnarray}}%

When this equation is treated as a partial differential equation for $B$, it can be linearized by a change of variable\cite{galehouse.ijtp.1,rankin.caqg,rankin.ijtp}. The substitution
\begin{equation}
B=\psi ^{-2}
\label{linear.change}
\end{equation}
produces
{\samepage 
\begin{eqnarray}
& (1/\sqrt {-\hat g}\, ) & (\sqrt {-\hat g}\, \hat g^{\mu \nu }
\psi _{,\nu })_{,\mu }+2\hat g^{\mu \nu }v_\mu \psi _{,\nu }
\nonumber \\
&  & +(1/\sqrt {-\hat g}\, )(\sqrt {-\hat g}\, \hat g^{\mu \nu }
v_\nu )_{,\mu }\psi +\hat g^{\mu \nu }v_\mu v_\nu \psi =
{\textstyle{1 \over 6}}\, (C-\hat R)\psi 
\label{sch.wave.eq}
\end{eqnarray}}%
Of course in reality, this is only one equation in a fully coupled set of equations, but the point being raised is whether forms are appearing that are similar to those of a familiar, existing theory, in this case, the theory of the relativistic Schr\"{o}dinger field\cite{morse.feshbach} (Klein-Gordon field) with a metric tensor 
$\hat g_{\mu \nu }$.

The above point is developed in an earlier paper\cite{rankin.caqg}. However, some results bear repeating here. The divergence of  equation (\ref{gi.maxwell.result}) gives
\begin{equation}
\hat g^{\mu \nu }\hat \beta _{||\mu ||\nu }=
2(\hat \beta \hat v^\mu )_{||\mu }
\label{consv.charge}
\end{equation}
Then if equation (\ref{gi.basic.identity}) is used on half of the total 
$\hat v^\mu _{\; \; ||\mu }$ term in equation (\ref{consv.charge}), it will give
{\samepage 
\begin{eqnarray}
& (1/\sqrt {-\hat g}\, ) & (\sqrt {-\hat g}\, \hat g^{\mu \nu }
\hat \beta _{,\nu })_{,\mu }+2\hat g^{\mu \nu }
(-\hat v_\mu )\hat \beta _{,\nu }
\nonumber \\
&  & +(1/\sqrt {-\hat g}\, )[\sqrt {-\hat g}\, \hat g^{\mu \nu }
(-\hat v_\nu )]_{,\mu }\hat \beta +
\hat g^{\mu \nu }(-\hat v_\mu )(-\hat v_\nu )\hat \beta =
{\textstyle{1 \over 6}}\, (C-\hat R)\hat \beta 
\label{gi.sch.wave.eq}
\end{eqnarray}}%
Define the {\it conjugate} wavefunction $\xi $ by
\begin{equation}
\xi = \psi ^{-1}\hat \beta 
\label{conj.psi}
\end{equation}
Equations (\ref{gi.sch.wave.eq}), (\ref{sch.wave.eq}), and (\ref{gi.maxwell.result}), used together with this definition, then give
{\samepage 
\begin{eqnarray}
& (1/\sqrt {-\hat g}\, ) & (\sqrt {-\hat g}\, \hat g^{\mu \nu }
\xi _{,\nu })_{,\mu }-2\hat g^{\mu \nu }v_\mu \xi _{,\nu }
\nonumber \\
&  & -(1/\sqrt {-\hat g}\, )(\sqrt {-\hat g}\, \hat g^{\mu \nu }
v_\nu )_{,\mu }\xi +\hat g^{\mu \nu }v_\mu v_\nu \xi =
{\textstyle{1 \over 6}}\, (C-\hat R)\xi 
\label{comp.wave.eq}
\end{eqnarray}}%
\begin{equation}
\hat \beta = \xi \psi 
\label{psistar.psi}
\end{equation}
and
\begin{equation}
2j^2 \hat p^{\mu \nu }_{\; \; \; \; ||\nu }=
-6\hat g^{\mu \nu }(\xi _{,\nu }\psi - \psi _{,\nu }\xi -2\xi \psi v_\nu )
\label{sch.maxwell.result}
\end{equation}
Thus in fact, $\xi $ does obey a conjugate Schr\"{o}dinger-like form (minus $v_\mu $), and equations (\ref{gi.einstein.result}), (\ref{sch.maxwell.result}), (\ref{sch.wave.eq}), and (\ref{comp.wave.eq}) form a system as they stand that is very similar to a coupled 
Einstein-Maxwell-Schr\"{o}dinger set of equations for 
$\hat g_{\mu \nu }$, $v_\mu $, $\xi $, and $\psi $. Indeed, the primary differences are the presence of manifestly general relativistic terms, the presence of second derivatives of $\xi \psi $ in the stress tensor (a remnant of the higher order derivatives in the action), and the fact that $\xi $ is {\it not} required to be the algebraic complex conjugate of $\psi $, denoted here by ${}^\# \psi $ (to avoid later confusion with the symbol for the tensor dual). Other points, such as $\psi $, $\xi $, and $v_\mu $ not being complex or imaginary here, are easily remedied if gauge invariant variables are used originally.

To underscore the points on the equation system, temporarily pin down 
$v_\mu $ by defining $v_\mu =[e/(\hbar c)]A_\mu $, $C$ by letting 
$C=-1$\cite{galehouse.minus.C}, and the scale factor by setting 
$b_0 =(6m_0 ^2 c^2 )/\hbar^2 $, where $m_0$ is some particle rest mass. Assume the dimensionless coordinates in use, the $x^\mu $, are those of a locally Lorentzian frame defined by the local vanishing of 
$\{ ^{\: \hat \mu }_{\nu \alpha } \} $, and by 
$\hat g_{\mu \nu }=\hat \eta _{\mu \nu }=diag(1,-1,-1,-1)$ locally (other definitions of a tangent frame are possible). Then introduce standard lab coordinates $\bar x^\mu =x^\mu /\sqrt {b_0 }$. {\it Assuming} the magnitude of $\hat R$ is negligible compared to $C$, and that the deviations of 
$\hat g_{\mu \nu }$ from Lorentzian locally are ignorable, then equation (\ref{sch.wave.eq}) becomes after this change of variable
{\samepage 
\begin{eqnarray}
& \hat \eta ^{\mu \nu }\psi _{,\mu ,\nu } & +2[e/(\hbar c)]
\hat \eta^{\mu \nu }A_\mu \psi _{,\nu }+[e/(\hbar c)]
(\hat \eta ^{\mu \nu }A_\nu )_{,\mu }\psi \nonumber \\
& & +[e/(\hbar c)]^2 \hat \eta^{\mu \nu }A_\mu A_\nu \psi =
-[(m_0 c)/\hbar]^2\psi 
\label{real.wave.eq}
\end{eqnarray}}%
where the derivatives and potentials are now evaluated using the 
$\bar x^\mu $. This immediately demonstrates that the scale factor, 
$b_0 ^{-1/2}$, provides the major entry for Planck's constant into the formalism. The scale factor's specific value appears to be an unrestricted but universal input parameter of the model. Furthermore, if $C$ has values other than $C=1$ or $C=-1$, the scale factor should be able to absorb the absolute value of C which differs from unity. Thus, the only non-trivial choices for $C$ for real variables are the cases $1$ and $-1$. The appropriate value for $j^2$ is $j^2 =2[\hbar/(ec)]^2 Gb_0 $.

But this still leaves a factor of $i=\sqrt{-1}$ missing from the potentials if an exact match with Schr\"{o}dinger theory is desired, although Wheeler discusses the possibility of truly dropping the $i$\cite{wheeler.physrev}. But if the $i$ is preferred, London's original, imaginary fine structure constant essentially reenters the geometry\cite{london.orig,adler.london}. An earlier paper examines such models\cite{rankin.caqg}, and shows that the present case should generalize to the action
{\samepage 
\begin{eqnarray}
I & = & \int {\{ (\hat R-2\sigma )-{\textstyle{1 \over 2}}\, 
j^2(\hat p_{\mu \nu }\hat p^{\mu \nu })+{\textstyle{1 \over 2}}
[\hat \beta (\hat R+6\hat v^\mu _{\; \; ||\mu }
+6\hat v^\mu \hat v_\mu }\nonumber \\
  &   & -e^{-i\hat \phi })+2\hat \alpha ^{\mu \nu }
\hat p_{\mu \nu }+CC]\} \sqrt {-\hat g}\, d^4x
\label{complex.gi.action}
\end{eqnarray}}%
Here, the action is considered to be a functional of purely real 
$\hat g^{\mu \nu }$ and $\hat \alpha ^{\mu \nu }$ (an antisymmetric tensor Lagrange multiplier), and the complex quantities $\hat \beta $ and $\hat v_\mu $. The $\hat \alpha ^{\mu \nu }$ term is a constraint that requires 
$\hat p_{\mu \nu }$ to be purely imaginary, the $CC$ denotes the complex conjugate of everything that precedes it in its containing brackets, and the 
$e^{-i\hat \phi }$ term is the generalization of $C$ to now include all points on the unit circle in the complex plane, where $\hat \phi $ is a constant. {\it This generalized action itself would be difficult to phrase without gauge invariant variables, since once the electromagnetic field becomes imaginary, complex gauge transformations are admissible.} Then only gauge invariant variables provide any sort of metric tensor which can unambiguously be required to be real.

The results of the variation of equation (\ref{complex.gi.action}) can be expressed as equations (\ref{gi.basic.identity}), (\ref{linear.change}), (\ref{sch.wave.eq}), (\ref{consv.charge}), (\ref{gi.sch.wave.eq}), (\ref{conj.psi}), (\ref{comp.wave.eq}), and (\ref{psistar.psi}), with
\begin{equation}
(\hat p_{\mu \nu }+CC)=0
\label{imag.maxwell}
\end{equation}
\begin{equation}
2j^2 \hat p^{\mu \nu }_{\; \; \; \; ||\nu }=
-3\hat g^{\mu \nu }(\xi _{,\nu }\psi - \psi _{,\nu }\xi 
-2\xi \psi v_\nu -CC)
\label{imag.sch.maxwell.result}
\end{equation}
{\samepage 
\begin{eqnarray}
\hat R_{\mu \nu }-{\textstyle{1 \over 2}}\, \hat R\hat g_{\mu \nu }+
\sigma \hat g_{\mu \nu } & = & -j^2(\hat p_\mu ^{\; \; \gamma }
\hat p_{\gamma \nu }+{\textstyle{1 \over 4}}\, \hat g_{\mu \nu }
\hat p_{\gamma \tau }\hat p^{\gamma \tau })-{\textstyle{1 \over 2}}\, 
\{ \xi \psi (\hat R_{\mu \nu }-{\textstyle{1 \over 2}}\, 
\hat R\hat g_{\mu \nu })\nonumber \\
 &   & -3[(\xi _{,\mu }-\xi v_\mu )(\psi _{,\nu }+\psi v_\nu )+
(\xi _{,\nu }-\xi v_\nu )(\psi _{,\mu }+\psi v_\mu )\nonumber \\
 &   & -(\xi _{,\gamma }-\xi v_\gamma )(\psi _{,\tau }+\psi v_\tau )
\hat g^{\gamma \tau }\hat g_{\mu \nu }
-{\textstyle{1 \over 6}}\, e^{-i\hat \phi }\xi \psi 
\hat g_{\mu \nu }]\nonumber \\
 &   & +[(\xi \psi )_{||\mu ||\nu }
-(\xi \psi )_{||\gamma ||\tau }\hat g^{\gamma \tau }
\hat g_{\mu \nu }]+CC\} 
\label{imag.sch.einstein.result}
\end{eqnarray}}%
and
\begin{equation}
\hat R-4\sigma ={\textstyle{1 \over 2}}\, (e^{-i\hat \phi }\xi \psi +CC)
\label{imag.rhat}
\end{equation}
The fact that equation (\ref{consv.charge}) is still true in this case is quite important in obtaining some of these results. It is preserved because the divergence of $\hat \alpha ^{\mu \nu }_{\; \; \; \; ||\nu }$ vanishes identically, as does the divergence of 
$\hat p^{\mu \nu }_{\; \; \; \; ||\nu }$\cite{rankin.caqg}.

Using the earlier $b_0 $ and $v_\mu =[(ie)/(\hbar c)]A_\mu $, these results of equation (\ref{complex.gi.action}) should correspond exactly to coupled Einstein-Maxwell-Schr\"{o}dinger theory for $\hat \phi =\pi $ and 
$j^2 =-2[\hbar/(ec)]^2 Gb_0 $ (note the minus), except for the second derivative terms in the stress tensor, the algebraic independence of $\xi $ and ${}^\# \psi $, and general relativistic terms which include additional factors of $\hat R_{\mu \nu }$ or $\hat R$. Moreover, as will be discussed below, the divergence of the second derivative terms in the stress tensor is 
$\hat R_\mu ^{\; \; \nu }(\xi \psi )_{,\nu }$, which limits them to a purely general relativistic interaction with the rest of the stress tensor. However, note that in all of these equations, {\it it is the gauge invariant metric that obeys the form recognizable as Einstein's equations, and which appears in the wave equations and Maxwell's equations.} The Weyl metric is determined in the solution, as it must be, but it emerges almost as an afterthought through equations (\ref{def.ghat}) and (\ref{linear.change}), which combined with $C=e^{-i\hat \phi }$ now give
\begin{equation}
g_{\mu \nu }=e^{-i\hat \phi }\psi ^2 \hat g_{\mu \nu }
\label{weyl.metric.solution}
\end{equation}
Note once again that {\it no} particular gauge has been fixed, since the value of $\psi $ will depend on whatever arbitrary electromagnetic gauge is imposed on $v_\mu $. Note also that in the model based on equation (\ref{complex.gi.action}), the Weyl metric may be complex, but the gauge invariant metric remains real.

To sum it up so far, the advantages of using gauge invariant variables are
\begin{itemize}
\item They often reduce the mathematical labor.
\item They may quickly isolate recognizable mathematical forms.
\item They should simplify specification of boundary conditions and correlation with physical observables.
\item They may aid in generalizing a model.
\end{itemize}
To these, I might add that they do these things without sacrificing the original gauge freedom of a structure. In fact, they may actually point to possible extensions of that freedom, as with equation (\ref{complex.gi.action}), which leads to a model allowing {\it complex} gauge transformations, yet with real observables (or completely imaginary ones, which work just as well)\cite{rankin.caqg}.

On another level, the use of gauge invariant variables justifies the results obtained in many earlier works in such geometries using the gauge $B=constant$\cite{weyl.stm,adler.london}. It should be obvious from equations (\ref{def.vhat}) and (\ref{def.ghat}) that the gauge invariant variables isolate essentially the same forms found in these earlier works, but {\it without} using any specific gauge. Thus, their use represents the formally correct method of obtaining the same simplifications without invoking gauge restrictions or losing generality, and from actions stated originally in the gauge invariant variables. As an example of this, the form equivalent to the original Weyl action\cite{weyl.stm} will be included below.

\section{Discussion of the Geometric Schr\"{o}dinger Behavior}
\label{quantum.discussion}

Since the action of equation (\ref{complex.gi.action}) is stated above to yield almost exactly the equations for a coupled 
Einstein-Maxwell-Schr\"{o}dinger classical field, this section will examine that result in more detail. First, the original Weyl case will be developed as a contrast. Then the above results will be contrasted with standard Einstein-Maxwell-Schr\"{o}dinger field theory in more detail. Finally, the remaining differences between the above theory and conventional 
Einstein-Maxwell-Schr\"{o}dinger field theory will be examined, along with the possible relationship to quantum mechanics of particles. This last issue actually will appear as the problem of discovering particle behavior at all in this continuum theory.

\subsection{Contrast with Weyl's Original Theory}

The original Weyl theory develops from the same type of geometry as the above, but from a different action\cite{weyl.stm,adler.london}. A contrast is most easily developed by rephrasing Weyl's action in terms of (real) gauge invariant variables\cite{rankin.caqg}. This gives
\begin{equation}
I=\int {[(\hat R-{\textstyle{1 \over 2}})-{\textstyle{1 \over 2}}\, 
j^2(\hat p_{\mu \nu }\hat p^{\mu \nu })+6\hat v^\mu \hat v_\mu +
\hat \beta (\hat R+6\hat v^\mu _{\; \; ||\mu }
+6\hat v^\mu \hat v_\mu -1)]\sqrt {-\hat g}\, d^4x}
\label{orig.gi.weyl.action}
\end{equation}

First note that the constraint term in the action is identical to the constraint in equation (\ref{complex.gi.action}) except for the specialization of $\hat \phi$ to zero. All actions in the gauge invariant variables will have the same form for the constraint, since it represents equation (\ref{gi.basic.identity}), a purely {\it kinematic} condition intrinsic to any such geometry. That condition by itself produces a Schr\"{o}dinger-like form (see equation (\ref{sch.wave.eq})) provided merely that the metric is approximately Lorentzian, and the magnitude of $\hat R$ is negligible compared to one. Thus, if Schr\"{o}dinger-like behavior is not to manifest itself in the geometry, the action must specify a dynamics which somehow masks it.

The field equations derived from equation (\ref{orig.gi.weyl.action}) are
{\samepage 
\begin{eqnarray}
\hat R_{\mu \nu }-{\textstyle{1 \over 2}}\, \hat R\hat g_{\mu \nu }+
{\textstyle{1 \over 4}}\, \hat g_{\mu \nu }
 & = & -j^2(\hat p_\mu ^{\; \; \gamma }\hat p_{\gamma \nu }+
{\textstyle{1 \over 4}}\, \hat g_{\mu \nu }
\hat p_{\gamma \tau }\hat p^{\gamma \tau })
\nonumber \\
 &   & -6\hat v_\mu \hat v_\nu
+3\hat v^\gamma \hat v_\gamma \hat g_{\mu \nu }
\label{gi.weyl.einstein.result}
\end{eqnarray}}%
\begin{equation}
2j^2 \hat p^{\mu \nu }_{\; \; \; \; ||\nu }=12\hat v^\mu 
\label{gi.weyl.maxwell.result}
\end{equation}
\begin{equation}
\hat \beta =0
\label{gi.weyl.betahat.result}
\end{equation}
and,
\begin{equation}
6\hat v^\mu \hat v_\mu =1-\hat R
\label{gi.weyl.identity}
\end{equation}

Thus ironically, the primary action used by Weyl does completely suppress the intrinsic, Schr\"{o}dinger-like form. Clearly $\hat v^\mu _{\; \; ||\mu }=0$ here, and the Schr\"{o}dinger-like form of equation (\ref{sch.wave.eq}) is replaced by equation (\ref{gi.weyl.identity}). This gives a classical, Hamilton-Jacobi form when the magnitude of $\hat R$ is negligible compared to one\cite{rankin.ijtp,landau.lifshitz}. {\it Any Schr\"{o}dinger tendencies are suppressed.} This in turn, illustrates yet another possible advantage to use of gauge invariant variables. {\it Simpler actions in gauge invariant variables, rather than the Weyl variables, may lead to more realistic models.} When the Weyl variables are used, the action of equation (\ref{example.main.action}) is clearly the more complicated appearing action when contrasted with Weyl's original action\cite{weyl.stm,adler.london}, which is simply
\begin{equation}
I=\int {[ {\textstyle{1 \over 2}}\, B^2 
-{\textstyle{1 \over 2}}\, j^2 p_{\mu \nu }p^{\mu \nu }]
\sqrt {-g}\, d^4 x}
\label{original.weyl.action}
\end{equation}

But besides being simpler in form than equation (\ref{orig.gi.weyl.action}) in terms of gauge invariant variables, the action of equation (\ref{complex.gi.action}) has another important attribute. Except for the imaginary value of the electromagnetic field, it is the obvious generalization to these geometries of the standard action for coupled 
Einstein-Maxwell fields with no continuous electromagnetic sources\cite{adler.london}. In fact, there is no particular reason other than convention that an imaginary electromagnetic field could not also have been used in the ``parent'' theory. Indeed, that ``free fields'' case (no continuous electromagnetic sources) is still an allowed result, since the allowed case $\xi = 0$ clearly gives the standard coupled Einstein-Maxwell equations with no continuous Maxwell sources\cite{rankin.caqg}. In contrast, the equations resulting from the action of equation (\ref{orig.gi.weyl.action}) {\it will not} admit this standard, coupled, free fields case, as was noticed at the time of Weyl's original work\cite{eddington.mtr}. Thus, except for the imaginary character of the electromagnetic field, the action of equation (\ref{complex.gi.action}) is actually the most immediately logical candidate for a direct transposition of Einstein-Maxwell theory to these geometries. When $\hat \beta =0$, the usual free field equations result (singular electromagnetic sources, or no electromagnetic sources), and equation (\ref{sch.wave.eq}) would seem then to have no obvious {\it physical} content, although it still completes the definition of the Weyl or Weyl-Cartan geometry. But when $\hat \beta \neq 0$, the full set of equations produced by equation (\ref{complex.gi.action}) indicates that electromagnetic sources are present, along with additional stress-energy terms. As will be detailed even further below, all these source terms have forms which are very close to those familiar from the theory of the Schr\"{o}dinger field. Thus, equations (\ref{sch.wave.eq}) and (\ref{comp.wave.eq}) are {\it not} just mathematically similar to the Schr\"{o}dinger equation and its conjugate, but actually can be tied to source motions (mechanics) closely resembling those familiar from Schr\"{o}dinger theory. {\it In these geometries, standard Einstein-Maxwell theory with a purely imaginary valued electromagnetic field, automatically produces a Schr\"{o}dinger-like, additional set of fields which form Schr\"{o}dinger-like, ``matter'' sources for the Maxwell and Einstein equations}. This is quite different from the original Weyl action's results, and is the second, main point of this paper. It provides the basic motivation for feeling these results may provide an important clue to the proper relationship between general relativistic, geometric theories, and theories of quantum mechanics. I feel this multiple equation similarity, involving all the field equations present, is unlikely to be simple coincidence.

However, this result is purchased at a price. In order not to lose the physical content (the source motion) of the result, {\it the electromagnetic sources must be taken as truly continuous}, and with values given literally by equation (\ref{imag.sch.maxwell.result}). Singular sources cause a reversion to the free field case, and the automatic, Schr\"{o}dinger-like source motion is lost. This forces the abandonment of the usual notion of the particle and of any probabilistic interpretation of the wavefunctions, if sources are assumed {\it not} to be singular in nature. In place of that, the formalism must now account for particle-like behavior from the classical field equations, or their offspring in some modified theory of this type, or perhaps a ``second quantized'' version of the field theory. As it stands now, the formalism portrays the sources as a single, continuous ``fluid'' of varying intensity, even if localized concentrations of the sources happen to appear, which might pass as individual ``particles''. This point will be revisited shortly.

\subsection{Contrast with Einstein-Maxwell-Schr\"{o}dinger Fields}

Unless otherwise noted, the gauge will be restricted hereafter to only those gauges which leave the electromagnetic potential, $v_\mu $, purely imaginary. Since $\hat p_{\mu \nu }$ is purely imaginary, this is hardly a restriction in practice. Indeed, all standard electromagnetic gauges fall inside this family, and gauges which do not do so would probably be regarded as bizarre by physicists, even when they are allowed. However, gauges obeying this restriction all have the convenient property that
\begin{equation}
v_\mu =[(ie)/(\hbar c)]A_\mu =-\, {}^\# v_\mu 
\label{conj.minus.vmu}
\end{equation}

Furthermore, assume $\hat \phi =\pi $, which gives $C=e^{-i\hat \phi }=-1$. This means that the complex conjugate of equation (\ref{comp.wave.eq}) for $\xi $ gives exactly the same form as equation (\ref{sch.wave.eq}), with the same electromagnetic potential, the same metric tensor, and the same $C$. This shows immediately that both $\psi $ and ${}^\# \xi $ are solutions to the same equation,
{\samepage 
\begin{eqnarray}
& (1/\sqrt {-\hat g}\, ) & (\sqrt {-\hat g}\, \hat g^{\mu \nu }
\psi _{,\nu })_{,\mu }+2\hat g^{\mu \nu }v_\mu 
\psi _{,\nu }\nonumber \\
&  & +(1/\sqrt {-\hat g}\, )(\sqrt {-\hat g}\, \hat g^{\mu \nu }
v_\nu )_{,\mu }\psi +\hat g^{\mu \nu }v_\mu v_\nu \psi 
=-{\textstyle{1 \over 6}}\, (1+\hat R)\psi 
\label{main.wave.eq}
\end{eqnarray}}%
where the scale factor choice of $b_0 =(6m_0 ^2 c^2 )/\hbar^2 $ is also still implied.

To indicate that both $\psi $ and ${}^\# \xi $ are solutions to equation (\ref{main.wave.eq}), define
\begin{equation}
\psi _1 =\psi 
\label{def.psi1}
\end{equation}
and
\begin{equation}
\psi _2 ={}^\# \xi 
\label{def.psi2}
\end{equation}
Equation (\ref{def.psi2}) means that
\begin{equation}
\xi ={}^\# \psi _2 
\label{xi.conj.psi2}
\end{equation}
This allows equations (\ref{imag.sch.maxwell.result}) through (\ref{imag.rhat}) to be rewritten as
\begin{equation}
2j^2 \hat p^{\mu \nu }_{\; \; \; \; ||\nu }=
-3\hat g^{\mu \nu }(\, {}^\# \psi _{2,\nu }\psi _1 - 
\psi _{1,\nu }\, {}^\# \psi _2 -2\, {}^\# \psi _2 \psi _1 v_\nu -CC)
\label{imag.sch.maxwell.result.2}
\end{equation}
{\samepage 
\begin{eqnarray}
\hat R_{\mu \nu }-{\textstyle{1 \over 2}}\, \hat R\hat g_{\mu \nu }+
\sigma \hat g_{\mu \nu } & = & -j^2(\hat p_\mu ^{\; \; \gamma }
\hat p_{\gamma \nu }+{\textstyle{1 \over 4}}\, \hat g_{\mu \nu }
\hat p_{\gamma \tau }\hat p^{\gamma \tau })-{\textstyle{1 \over 2}}\, 
\{ \, {}^\# \psi _2 \psi _1 (\hat R_{\mu \nu }-{\textstyle{1 \over 2}}\, 
\hat R\hat g_{\mu \nu })\nonumber \\
 &   & -3[\, {}^\# (\psi _{2,\mu }+\psi _2 v_\mu )
(\psi _{1,\nu }+\psi _1 v_\nu )+\, {}^\# (\psi _{2,\nu }+\psi _2 
v_\nu )(\psi _{1,\mu }+\psi _1 v_\mu )\nonumber \\
 &   & -\, {}^\# (\psi _{2,\gamma }+\psi _2 v_\gamma )
(\psi _{1,\tau }+\psi _1 v_\tau )\hat g^{\gamma \tau }
\hat g_{\mu \nu }+{\textstyle{1 \over 6}}\, {}^\# 
\psi _2 \psi _1 \hat g_{\mu \nu }]\nonumber \\
 &   & +[(\, {}^\# \psi _2 \psi _1 )_{||\mu ||\nu }
-(\, {}^\# \psi _2 \psi _1 )_{||\gamma ||\tau }\hat g^{\gamma \tau }
\hat g_{\mu \nu }]+CC\} 
\label{imag.sch.einstein.result.2}
\end{eqnarray}}%
and
\begin{equation}
\hat R-4\sigma =-{\textstyle{1 \over 2}}\, (\, {}^\# \psi _2 \psi _1 +CC)
\label{imag.rhat.2}
\end{equation}
where the $\#$ in front of parentheses operates only on those terms inside the nearest set of parentheses. These forms begin to highlight both the similarities, and the differences between this model, and the standard theory of the coupled Einstein-Maxwell-Schr\"{o}dinger fields (see the appendix).

One immediate difference with the standard theory is that wavefunctions are {\it not} normalized to have magnitudes which are square integrable to unity. Instead, based on equation (\ref{imag.rhat.2}), the product 
${}^\# \psi _2 \psi _1$ is seen to be of the same magnitude as $\hat R$, which would normally be a very small quantity (exact relations can be derived between the dimensionless wavefunctions of this paper, and ``standard'' wavefunctions of textbooks, by following the appendix, and comparing it with the equations below for $\zeta _n $). Furthermore, for the $b_0 $ specified,
\begin{equation}
j^2 =-12Gm_0 ^2 /e^2 =-2.88\cdot 10^{-42}
\label{val.j2}
\end{equation}
where $m_0 $ has been chosen as the electron rest mass (this will be examined shortly). Thus, $j^2 $ is a very small quantity, so that wavefunctions with very small magnitudes are consistent with reasonable magnitudes of electromagnetic sources in equation (\ref{imag.sch.maxwell.result.2}).

But the most obvious difference with standard theory is the presence of both $\psi _1 $ and $\psi _ 2$ as possibly separate solutions to equation (\ref{main.wave.eq}) in the equations. To clarify this, define new wavefunctions $\zeta _1 $ and $\zeta _2 $ by\cite{wheeler.trans}
\begin{equation}
\psi _1 =(1/\sqrt{2}\, )(\zeta _1 +\zeta _2 )
\label{def.zetas.1}
\end{equation}
and
\begin{equation}
\psi _2 =(1/\sqrt{2}\, )(\zeta _1 -\zeta _2 )
\label{def.zetas.2}
\end{equation}
These new wavefunctions will uncouple the wavefunction terms in the equations almost entirely. Equations (\ref{imag.sch.maxwell.result.2}), (\ref{imag.sch.einstein.result.2}), and (\ref{imag.rhat.2}) become
\begin{equation}
2j^2 \hat p^{\mu \nu }_{\; \; \; \; ||\nu }=
3\sum _{n=1}^2 (-1)^n \hat g^{\mu \nu }(\, {}^\# \zeta _{n,\nu }\zeta _n 
- \zeta _{n,\nu }\, {}^\# \zeta _n -2\, {}^\# \zeta _n \zeta _n v_\nu )
\label{zeta.sch.maxwell.result}
\end{equation}
{\samepage 
\begin{eqnarray}
\hat R_{\mu \nu }-{\textstyle{1 \over 2}}\, \hat R\hat g_{\mu \nu }+
\sigma \hat g_{\mu \nu } & = & -j^2(\hat p_\mu ^{\; \; \gamma }
\hat p_{\gamma \nu }+{\textstyle{1 \over 4}}\, \hat g_{\mu \nu }
\hat p_{\gamma \tau }\hat p^{\gamma \tau })+{\textstyle{1 \over 2}}\, 
\sum _{n=1}^2 (-1)^n \{ \, {}^\# \zeta _n \zeta _n (\hat R_{\mu \nu }-
{\textstyle{1 \over 2}}\, \hat R\hat g_{\mu \nu })\nonumber \\
 &   & -3[\, {}^\# (\zeta _{n,\mu }+\zeta _n v_\mu )
(\zeta _{n,\nu }+\zeta _n v_\nu )+\, {}^\# (\zeta _{n,\nu }+\zeta _n 
v_\nu )(\zeta _{n,\mu }+\zeta _n v_\mu )\nonumber \\
 &   & -\, {}^\# (\zeta _{n,\gamma }+\zeta _n v_\gamma )
(\zeta _{n,\tau }+\zeta _n v_\tau )\hat g^{\gamma \tau }
\hat g_{\mu \nu }+{\textstyle{1 \over 6}}\, {}^\# 
\zeta _n \zeta _n \hat g_{\mu \nu }]\nonumber \\
 &   & +[(\, {}^\# \zeta _n \zeta _n )_{||\mu ||\nu }
-(\, {}^\# \zeta _n \zeta _n )_{||\gamma ||\tau }\hat g^{\gamma \tau }
\hat g_{\mu \nu }]\} 
\label{zeta.sch.einstein.result}
\end{eqnarray}}%
and
\begin{equation}
\hat R-4\sigma =-{\textstyle{1 \over 2}}\, (\, {}^\# \zeta _1 \zeta _1 
-\, {}^\# \zeta _2 \zeta _2 )
\label{zeta.rhat}
\end{equation}
Furthermore, since both $\psi _1 $ and $\psi _2 $ are solutions to equation (\ref{main.wave.eq}), one can add and subtract versions of equation (\ref{main.wave.eq}) to get
{\samepage 
\begin{eqnarray}
& (1/\sqrt {-\hat g}\, ) & (\sqrt {-\hat g}\, \hat g^{\mu \nu }
\zeta _{n,\nu })_{,\mu }+2\hat g^{\mu \nu }v_\mu 
\zeta _{n,\nu }+(1/\sqrt {-\hat g}\, )(\sqrt {-\hat g}\, \hat g^{\mu \nu }
v_\nu )_{,\mu }\zeta _n \nonumber \\
&  & +\hat g^{\mu \nu }v_\mu v_\nu \zeta _n 
-{\textstyle{1 \over 12}}\, (\, {}^\# \zeta _1 \zeta _1 
-\, {}^\# \zeta _2 \zeta _2 )\zeta _n 
=-{\textstyle{1 \over 6}}\, (1+4\sigma )\zeta _n 
\label{zeta.nonlin.wave.eq}
\end{eqnarray}}%
for either value of $n$. Equation (\ref{zeta.rhat}) is used in obtaining this result.

However, equations (\ref{zeta.sch.maxwell.result}), (\ref{zeta.sch.einstein.result}), (\ref{zeta.rhat}), and (\ref{zeta.nonlin.wave.eq}) are still the result of the action of equation (\ref{complex.gi.action}), subject only to the reasonable gauge restriction that $v_\mu $ is purely imaginary. But these same equations can also be derived from a ``standard form'' action for a {\it pair} of Klein-Gordon (relativistic Schr\"{o}dinger) fields coupled to the Einstein-Maxwell fields via standard couplings (see the appendix). Thus, the unorthodox action of equation (\ref{complex.gi.action}) is equivalent to the ``standard form'' action of equation (\ref{dual.conf.kg.action}) insofar as it produces {\it the same equations of motion for the fields}. If the further restriction is made that $\zeta _1 =0$, then the remaining fields obey ``standard form'' field equations for a single Klein-Gordon field coupled to the 
Einstein-Maxwell fields in the conventional manner\cite{wald}.

On the other hand, there is no a priori reason to restrict $\zeta _1 $ to zero, even though it should always be possible to require this with reasonable boundary conditions. This leads to the second, basic cost of generating automatic Schr\"{o}dinger behavior from the standard, 
Einstein-Maxwell action in a Weyl-like geometry. {\it An auxiliary, Schr\"{o}dinger field enters into the coupled equations, in addition to the ``normal'' Schr\"{o}dinger field}. This auxiliary field enters into equations (\ref{zeta.sch.maxwell.result}) and (\ref{zeta.sch.einstein.result}) through terms with reversed sign.

However, in spite of this reversed sign aspect of $\zeta _1 $, it is not clear that it is always desirable to have $\zeta _1 =0$. For example, the case ${}^\# \zeta _1 =\zeta _1 $, ${}^\# \zeta _2 =-\zeta _2 $, and $v_\mu =0$, leads to an entire family of possible solutions in which there are no electromagnetic fields or sources, and the resulting neutral ``matter'' fields involve only curvature related quantities (Weyl or Riemannian) and the metric. Some of these cases have been examined by Frank Taylor and myself in a cosmological context\cite{taylor.rankin}, and for a slightly more general formalism\cite{rankin.caqg}. When the formalism allows dynamic changes in the sign of the ``rest mass'' squared term in the equation for $\zeta _n $, results include both inflationary, and ``Big Bang'' type solutions. No smooth transition between such cases has yet been found. Nevertheless, this illustrates how some cases may be of interest in which $\zeta _1 $ is not zero.

\subsection{General Relativistic Effects}

Even if $\zeta _1 =0$, equation (\ref{zeta.sch.einstein.result}) will still contain terms not found in the standard, Einstein-Maxwell-Schr\"{o}dinger field theory (see the appendix). The terms 
${}^\# \zeta _2 \zeta _2 (\hat R_{\mu \nu }-{\textstyle{1 \over 2}}\, 
\hat R\hat g_{\mu \nu })$ are part of the difference, and are clearly ``general relativistic'' in character since they contain $\hat R_{\mu \nu }$. As manifestly general relativistic terms, they can be expected not to create major differences with the standard theory unless the gravitational field has components approaching unity in magnitude. In other words, they are higher order effects.

On the other hand, the terms $(\, {}^\# \zeta _2 \zeta _2 )_{||\mu ||\nu }
-(\, {}^\# \zeta _2 \zeta _2 )_{||\gamma ||\tau }\hat g^{\gamma \tau }
\hat g_{\mu \nu }$ appear likely to have magnitudes comparable with the standard terms not containing $v_\mu $ in the stress tensor, and are of greater concern when considering differences with the standard theory. But {\it any} terms in the stress tensor must have some physical consequence in order to be detected. In the case of these terms, the two possible consequences are the change in the gravitational field produced, and their interaction with the rest of the stress tensor through their divergence. But the first effect is technically still a general relativistic effect, and would be ignored in {\it special relativistic}, Maxwell-Schr\"{o}dinger field theory. And, the divergence of these terms is easily shown to be 
$\hat R_\mu ^{\; \; \nu }(\, {}^\# \zeta _2 \zeta _2 )_{,\nu }$, which contains $\hat R_{\mu \nu }$ explicitly, and is clearly a higher order effect itself. Thus, these terms also produce only general relativistic modifications to standard theory. In this case, they could be expected to show up as an apparent, nonconservation of energy in the standard theory under unusual circumstances which permit components of $\hat R_{\mu \nu }$ to approach unit magnitude, or via their effect on the overall gravitational field strength. Otherwise, they are essentially undetectable, since they are independently conserved in the flat spacetime limit.

Thus in all of these terms, it is the approach toward unit magnitude of at least some components of $\hat R_{\mu \nu }$ which appears important. Without this, only the standard terms are significant throughout the equations. Therefore, to get a crude estimate of just when this might occur, assume that at least some solutions to these equations can be approximated crudely by the Reissner-N\"{o}rdstrom solution for a general relativistic, electric monopole\cite{adler.london}, a fact not really known at this point. In that solution, assume that the pertinent charge is the electronic charge magnitude, $e$, and that the scale factor applicable is still 
$b_0 =(6m_0 ^2 c^2 )/\hbar^2 $. This last point is quite important, and therefore open to challenge, but perhaps the best current justification for it is simply that without it, there is little hope that equation (\ref{main.wave.eq}) can be brought into any agreement with atomic scale phenomena. Presumably, such phenomena would involve smeared out, electronic charge distributions with little opportunity to generate any strong gravitational field necessary to cause $\hat R$ to contribute significantly to the ``rest mass'' term in equation (\ref{main.wave.eq}), assuming the behavior of $\hat R$ is ever suitable to produce that effect. While extensions to this formalism may alter the situation there somewhat\cite{rankin.mg7,rankin.caqg}, this result argues for the current choice of scale factor here. Furthermore, the scale factor seems inflexible once set, since a change to its value represents a {\it global} scale transformation. Only a truly isolated system, if such exists, can be envisioned as having some different choice of scale factor from the rest of the universe without inconsistencies. The scale factor should be envisioned as setting the basic unit of measure in terms of more common, laboratory units.

With these assumptions, one finds that to lowest order, the largest components of
\begin{equation}
\hat R_{\mu \nu }\sim 12[(Gm_0 ^2 )/(\hbar c)][e^2 /(\hbar c)](1/r^4 )
\label{r.n.approx}
\end{equation}
and that this approaches unity for a dimensionless radius 
$r_0 =3.5\cdot 10^{-12}$, or in lab units, $5.6\cdot 10^{-23}\; cm.$. Any corresponding interaction cross section should contain the factor 
$\pi r_0 ^2$\cite{schiff}, so this crude estimate gives a very approximate cross section of $\sim 10^{-44}\; cm.^2 $, a very small effect indeed. While the numbers involved are below the current threshold of resolution for the pointlike nature of elementary charges (about $10^{-17}\; cm.$), the cross section is actually comparable with the low end of observed neutrino cross sections in matter\cite{texas92}. Thus, if the scale factor choice and other assumptions are reasonable ones, some predicted general relativistic effects in this model would be very small, but within range of existing experiments.

\subsection{Positivity of Energy and Related Issues}

The above indicates that for $\zeta _1 =0$, the additional terms in these equations introduce no deviations from conventional Maxwell-Schr\"{o}dinger field theory in the special relativistic limit. However, the case 
$\zeta _1 \neq 0$ is a different matter. The negative energy density of the $\zeta _1 $ portion of the stress tensor raises concern about appearance of gravitational repulsion. Such behavior is considered unphysical, since no case of it has ever been observed in nature. Additionally, the progenitor of this theory came from the action of equation (\ref{example.main.action}), a higher derivative action. Such actions are known to lead to classical field theories with a number of possible problems\cite{simon.physrev,rainer.qc}.

First note that the sign of the stress tensor terms has nothing to do with the symmetry between matter and antimatter. The mapping 
$\, {}^\# \zeta _n \rightarrow \zeta _n $, 
$\zeta _n \rightarrow \, {}^\# \zeta _n $, 
$\hat g_{\mu \nu }\rightarrow \hat g_{\mu \nu }$, and 
$v_\mu \rightarrow -v_\mu =\, {}^\# v_\mu $, clearly maps a given solution of equations (\ref{zeta.sch.maxwell.result}), (\ref{zeta.sch.einstein.result}), (\ref{zeta.rhat}), and (\ref{zeta.nonlin.wave.eq}) into another valid solution of these same equations. This is the matter-antimatter symmetry, and it has {\it no} effect on the sign of terms in the stress tensor.

Conversely, the mapping $\zeta _1 \rightarrow \zeta _2 $, 
$\zeta _2 \rightarrow \zeta _1 $ is not generally associated with any convenient mapping of any other field quantities which will produce a new solution from an existing one. For example, this mapping obviously tends to reverse the sign of the right side of equation (\ref{zeta.sch.maxwell.result}). However, if one then maps 
$v_\mu \rightarrow -v_\mu $ to try to bring the electromagnetic field into agreement with this, the $v _\mu $ terms on the right side of equation (\ref{zeta.sch.maxwell.result}) then reverse sign again. This leaves them with the wrong sign to restore the proper values for a solution. The nonlinear terms in equation (\ref{zeta.nonlin.wave.eq}), and the metric determined by equation (\ref{zeta.sch.einstein.result}) provide additional disruption to the simple generation of a new solution by such methods. While these points do not rule out antigravitating solutions, they do mean such solutions need not have the same properties as gravitating solutions. The key point involved in this argument is that the gravitational and electromagnetic {\it self fields} of the $\zeta _n $ must be included in the overall solution. In other words, any solutions must be fully self consistent.

Thus, while no proof currently exists that rules out antigravitating solutions, there is some hope that such solutions might be foreign or unobservable in our environment. Possibilities range from the failure of such ``matter'' to coalesce into stars and planets, to possible lack of stability of such solutions, or swamping of any antigravitational effects by more common, or more massive gravitating solutions. In any event, Tolman indicates that a satisfactory condition for gravitation at large distances from a quasi-static, localized source in a space-time which approaches Lorentzian at spatial infinity, is $U > 0$, where
\begin{equation}
U=\int {[\hat G_\mu ^\mu -2\hat G_0 ^0 ]\sqrt {-g}\, d^3 x}
\label{tolman.integral}
\end{equation}
and $\hat G_\mu ^\nu =\hat R_\mu ^\nu -{\textstyle{1 \over 2}}\, %
\hat R\delta_\mu ^\nu $, the usual Einstein tensor\cite{tolman.rtc}. This condition may be as simple to grasp as any other indicator of asymptotic gravitation.

However, the possibility of antigravitating solutions remains a serious issue. If such solutions are present at all in our environment, even as a minor portion of it, current experimental evidence suggests that they would necessarily have negative passive and inertial mass densities, along with their negative, active gravitational mass density\cite{cook.will}. They could place this model into disagreement with observation in the future. I currently see no simple way to exclude this risk, since it seems basic to the formalism, and the automatic production of the Schr\"{o}dinger field forms from the combination of Einstein-Maxwell theory, and a Weyl-like geometry. On the other hand, this aspect of the equations may have important cosmological implications through production of inflation under certain, unusual conditions, primarily in an extended formalism\cite{taylor.rankin}. Thus, this additional freedom in the equations may also be beneficial in cases. In this regard, the study of such cosmological models has produced one example of an open universe in unstable equilibrium, with constant, negative energy density\cite{taylor.rankin}. Any disturbance to the equilibrium has been found to produce one of two results. Either the universe begins to expand, and its negative energy density very rapidly approaches zero (the expansion becomes curvature dominated), or the energy density quickly becomes positive as the universe contracts to a singular ``crunch''. Either way, the presence of a significant, negative energy density appears to be difficult to maintain, at least in one example.

At the same time, some of the other problems with higher derivative theories\cite{simon.physrev,rainer.qc} seem to be controlled in this model. When $\hat \phi =\pi $, the ``rest mass'' squared term in equation (\ref{main.wave.eq}) enters with the ``correct'' sign, so ``rest mass'' is real, and flat space time has  been found to be stable against perturbations\cite{taylor.rankin}, at least for a Friedmann-Robertson-Walker (FRW) metric. But as Simon notes\cite{simon.physrev}, this can also prevent inflation. It appears possible to remove this last limitation in a controlled, dynamic manner by extending the model with an antisymmetric part to the metric\cite{rankin.caqg}. This will be summarized briefly later.

\subsection{The Problem of Particle Existence and Quantum Behavior}

As noted above, particles can no longer simply be introduced as independent postulates into this formalism. This is one price of automatic production of Schr\"{o}dinger field forms purely from Einstein-Maxwell theory imbedded in a Weyl-like geometry. However, the particle properties of matter are a well known fact of our experience. There appear to be at least two possible responses to this dilemma. Each has been a conventional approach to this problem for physics at some period in its history.

The approach currently favored is the quantum field theoretic approach\cite{guidry}. The field equations derived from a Lagrangian are subjected to ``second quantization'' via standard rules for field quantization. However, this encounters problems here on two different levels. On a practical level, the true independent field variables in the theory are either the Weyl metric and four vector, or the gauge invariant metric, four vector, and Lagrange multiplier $\hat \beta $. These choices can be also enhanced by including the affine connection forms in the procedure. But either way, the task appears unlikely to be any simpler than the quantization of the gravitational field, a problem rife with well known difficulties\cite{penrose}. Nevertheless, this would currently be the conventional way to approach this problem.

However on a philosophic level, it seems odd to arbitrarily impose second quantization onto a theory in which the Schr\"{o}dinger field arises in such an entirely different fashion from its usual origin in first quantization. If this different approach provides any real clue to the relation between quantum theory and general relativity, then it seems more natural to search for a similar clue to ``second quantization''. Since second quantization introduces the concept of field quanta with particle characteristics\cite{guidry}, this is really a search for particle properties in the theory.

Prior to the ascendancy of quantum field theory, it was common to search for models of the electron and other particles in which they were constructed from a charge density, possibly with a divergent value at some point, and resulting from classical field equations, with or without ``Poincare stresses'' added\cite{jackson.self.force,sommerfeld,weyl.stm,eddington.mtr,%
einstein.particles,schrodinger,pauli}. Some of these efforts, particularly those of Einstein, aimed at a genuine, classical continuum model of the field sources. This model was to be devoid of singularities, and drew no sharp divisions between separate ``particles'', which were to be simply {\it localized concentrations} of the sources\cite{einstein.meaning,hoffmann,pauli}. While this may now seem foreign as an approach, it seems to me to be better suited to the model of this paper than an arbitrary attempt at field quantization. Certainly, the abandonment of the a priori particle concept mentioned earlier forces a true continuum picture of field sources onto the model. It is also worth noting that none of the earlier classical efforts had the possible advantage of intrinsic Schr\"{o}dinger-like, source behavior built-in to the field equations which the model of this paper has, although Weyl's original theory clearly came close to this, suppressing it only through an ``unfortunate'' choice for the action integral.

Along these lines, note that {\it if} such an explanation of ``particles'' could be successfully carried through, it need only produce particle-like behavior in cases in which the particle nature of matter manifests itself experimentally. In fact, to avoid problems with the negative experimental results currently associated with tests of hidden variables\cite{j.s.bell}, it might even be {\it necessary} that the particle nature actually be {\it produced} in the measuring process, at least to some degree. This approach would then attribute the quantum mechanics of such ``particles'' literally to the classical field equations (\ref{zeta.sch.maxwell.result}), (\ref{zeta.sch.einstein.result}), and (\ref{zeta.nonlin.wave.eq}), or their counterparts in an extended formalism. While it is currently speculative to presume this could be achieved, it is not speculation to note that this is the type of program which must succeed if this approach is valid (Pauli explains this particularly clearly\cite{pauli}). The alternatives appear to be either the recourse to second quantization noted above, or the abandonment of the idea that the automatic Schr\"{o}dinger behavior of this geometrically unified model is significant. Thus, this issue cannot simply be ignored.

Now since the current model resembles existing 
Einstein-Maxwell-Schr\"{o}dinger field theory closely, it might seem unlikely to provide any really new insights into the ``particle'' problem at this classical level. However, at least one major difference with the conventional theory has been noted above, the possibility of negative energy density matter terms in the stress tensor. Other differences exist as well at strong field strengths, such as the nonlinear terms in equation (\ref{zeta.nonlin.wave.eq}), which have been associated with the appearance of solitons\cite{solitons}. Thus, to explore a bit further, rewrite equation (\ref{zeta.sch.einstein.result}) as
{\samepage 
\begin{eqnarray}
\hat R_{\mu \nu }-{\textstyle{1 \over 2}}\, \hat R\hat g_{\mu \nu }
 & = & -[1+{\textstyle{1 \over 2}}\, (\, {}^\# \zeta _1 \zeta _1 -\, 
{}^\# \zeta _2 \zeta _2 )]^{-1}\{ \sigma \hat g_{\mu \nu } +j^2
(\hat p_\mu ^{\; \; \gamma }\hat p_{\gamma \nu }+{\textstyle{1 \over 4}}
\, \hat g_{\mu \nu }\hat p_{\gamma \tau }\hat p^{\gamma \tau })
\} \nonumber \\
 &   & -{\textstyle{1 \over 2}}\, [1+{\textstyle{1 \over 2}}\, 
(\, {}^\# \zeta _1 \zeta _1 -\, {}^\# \zeta _2 \zeta _2 )]^{-1}
\sum _{n=1} ^2 (-1)^n \{ 3[\, {}^\# (\zeta _{n,\mu }+\zeta _n v_\mu )
(\zeta _{n,\nu }+\zeta _n v_\nu )\nonumber \\
 &   & +\, {}^\# (\zeta _{n,\nu }+\zeta _n 
v_\nu )(\zeta _{n,\mu }+\zeta _n v_\mu )
-\, {}^\# (\zeta _{n,\gamma }+\zeta _n v_\gamma )
(\zeta _{n,\tau }+\zeta _n v_\tau )\hat g^{\gamma \tau }
\hat g_{\mu \nu }\nonumber \\
 &   & +{\textstyle{1 \over 6}}\, {}^\# 
\zeta _n \zeta _n \hat g_{\mu \nu }]
-[(\, {}^\# \zeta _n \zeta _n )_{||\mu ||\nu }
-(\, {}^\# \zeta _n \zeta _n )_{||\gamma ||\tau }\hat g^{\gamma \tau }
\hat g_{\mu \nu }]\} 
\label{zeta.sch.einstein.result.3}
\end{eqnarray}}%
Equations (\ref{zeta.sch.einstein.result.3}), (\ref{zeta.sch.maxwell.result}), and (\ref{zeta.nonlin.wave.eq}) are now the basic equations of the theory, along with 
$\hat p_{\mu \nu }=v_{\nu ,\mu }-v_{\mu ,\nu }$. The scale factor $b_0 $, 
$j^2 $, and the potential $v_\mu $ remain as previously defined.

In this form, equation (\ref{zeta.sch.einstein.result.3}) reveals another potentially important, new feature. The traditional stress-energy tensor has been multiplied throughout by the factor $[1+{\textstyle{1 \over 2}}\, %
(\, {}^\# \zeta _1 \zeta _1 -\, {}^\# \zeta _2 \zeta _2 )]^{-1}$. Equation (\ref{zeta.rhat}) indicates that for cases giving $\hat R > 4\sigma $, this factor will diverge if $\hat R \rightarrow 1+4\sigma $. The standard 
stress-energy terms could thus be enhanced arbitrarily by this scalar curvature ``feedback'' effect in strong matter fields (without any failure of conservation laws). Or, for cases in which $\hat R<4\sigma $, this same factor will tend to decrease the effect of the ``usual'' stress tensor terms. Either way, only the trace of the usual stress-energy would contribute to this. On the other hand, the terms in the electromagnetic stress tensor are not composed directly from $\zeta _1 $ or $\zeta _2 $, so there is no particular reason to expect that part of the stress-energy to approach zero if 
$\hat R \rightarrow 1+4\sigma $. The net result is the {\it possibility} of strong gravitational fields in the domain of elementary charges, with magnitudes that could differ considerably from the usual estimates based on vacuum solutions to the Einstein-Maxwell equations\cite{adler.london} where this feedback does not occur. The feedback is subtle however. For example, the simple, neutral matter, FRW cosmology mentioned previously\cite{taylor.rankin}, produces a stress tensor which conspires to remain smoothly finite as it passes through the limit 
$\hat R \rightarrow 1+4\sigma $ on the way to or from a ``Big Bang'' singularity, unless the time derivative of $(\zeta _1 ^2 +\zeta _2 ^2 )$ also approaches zero at the same limit time. This can be analytically demonstrated for this case. However, note that this case also contains no electromagnetic stress-energy.

Thus, equations (\ref{zeta.sch.einstein.result.3}), (\ref{zeta.sch.maxwell.result}), and (\ref{zeta.nonlin.wave.eq}) do contain new features which might effect the ``particle'' problem at the classical level. However, little that is certain is known at this time, and the same features might also eventually prove to be liabilities. The solutions for central symmetry should help answer this, but that case produces equations which have so far resisted analytical solution, and no numerical analysis has been attempted yet. The problem of ``particles'' thus remains open. If the above equations, or extensions with additional interactions (see below) fail to produce positive indications of some particle behavior, then this entire approach would appear less promising.

One additional comment may be appropriate. If this model is pursued as a classical field theoretic alternative to second quantization, and not just as a classical approximation, then it should probably also be contrasted with quantum field theories, and not just classical ones. In particular, its potential oddities might be viewed accordingly, such as possible negative energy density fields. Quantum field theories and some of their semiclassical approximations are known to produce negative energy densities under some circumstances\cite{ford}. The possible appearance of similar behavior here does not seem that unusual in this context.

\section{Brief Survey of Extended Models}
\label{brief.survey}

Some of the older searches for classical field models of particles never expected an answer solely from a theory of gravitation and electromagnetism\cite{jackson.self.force}. Furthermore, modern physics recognizes two additional interactions, as well as intrinsic spin\cite{guidry}, none of which is obvious in the model proposed to this point. The above equations also admit only a single value for the scale factor, which in turn restricts the quantity which acts like rest mass in equation (\ref{zeta.nonlin.wave.eq}) to a single value. All of these points call for some extension to the theory.

One natural choice for an extended formalism involves addition of a self dual, antisymmetric part to the metric tensor in the Weyl-Cartan geometry\cite{rankin.caqg}. While such an extension can also be envisioned for the pure Weyl case, it fits the already asymmetric Cartan case more naturally. Details are found in the earlier paper\cite{rankin.caqg}, but a brief introduction will be given here, primarily to show how this extension may itself be extended. While none of these extensions has shown the simple unity of the model presented above, there is limited positive correlation with Dirac theory\cite{rankin.caqg}. Additionally, the entire notion of an antisymmetric part to the metric appears likely to link up with older notions of intrinsic spin in Cartan geometries, and to provide a badly needed ability to adjust the ``rest mass'' parameter in the analog of equation (\ref{zeta.nonlin.wave.eq}).

\subsection{A Very Simple Attempt at Spin}

As noted, assume that the Weyl-Cartan geometry is the only one under consideration. If the metric is allowed to have an antisymmetric part as well as a symmetric part, and if the antisymmetric part is proportional to its own dual (self dual), the analog of equation (\ref{gi.basic.identity}) acquires a 
(non-spinor) term closely resembling the spin term in the second order Dirac equation\cite{rankin.caqg}. Such a restriction, that the antisymmetric part to the metric is self dual, prevents any path dependent changes in the length of any vectors under any kind of parallel transport. The self dual quality is also suggested by the fact that the Dirac spin tensor-spinor is self dual in its tensor indices. Since purely classical spinors, such as the eigenspinors of the electromagnetic field, can often be used to rephrase spinor equations as scalar equations\cite{rankin.caqg} or vice-versa, the non-spinor nature of the result need not be a {\it fundamental} problem. Such a self dual antisymmetric part to the metric also simplifies many of the complications associated with asymmetric metrics, although it requires admission of complex numbers. But these are already present in equation (\ref{complex.gi.action}).

Now the entire metric becomes $\hat m_{\mu \nu }$, the symmetric part is the real quantity $\hat g_{\mu \nu }$, and the antisymmetric part is 
$\hat a_{\mu \nu }=\hat A_{\mu \nu }+i\: {}^* \! \! \hat A_{\mu \nu }$, where the $*$ denotes the dual of the tensor following it. The action can be taken to be a functional of real $\hat g^{\mu \nu}$ and 
%
%
$\hat \alpha ^{\mu \nu }$, real, imaginary, or even complex 
$\hat A_{\mu \nu }$, and complex $\hat v_\mu $, $\hat \beta $, and 
$\hat \gamma $, where $\hat \gamma $ is another Lagrange multiplier. The gauge restrictions of much of section \ref{quantum.discussion} are dropped, and the action is given the form
{\samepage 
\begin{eqnarray}
I & = & \int {\{ (\hat R-2\sigma )-{\textstyle{1 \over 2}}\, 
j^2(\hat p_{\mu \nu }\hat p^{\mu \nu })+{\textstyle{1 \over 2}}
[\hat \beta (\hat R+6\hat v^\mu _{\; \; ||\mu }
+6\hat v^\mu \hat v_\mu +\hat a^{\mu \nu }\hat p_{\mu \nu }}
\nonumber \\
  &   & -e^{-i\hat \phi }-{\textstyle{1 \over 4}}\, 
e^{-i\hat \phi }\hat a)+2\hat \alpha ^{\mu \nu }
\hat p_{\mu \nu }+\hat \gamma \hat f(\hat a)+CC]\} 
\sqrt {-\hat g}\, d^4x
\label{complex.gi.ahat.action}
\end{eqnarray}}%
where $\hat a = \hat a_{\mu \nu }\hat a^{\mu \nu }$, $\hat f(\hat a)$ is a function of $\hat a$, and this action is a slight generalization of the earlier paper's action\cite{rankin.caqg}. Besides the obvious 
$e^{-i\hat \phi }$ factor, the main difference with the earlier paper is the replacement of the earlier constraint $\hat a=K^2$, where $K$ is constant, by the slightly more general constraint $\hat f(\hat a)=0$ instead (require 
$\hat f^\prime (\hat a)\neq 0$ at the roots of $\hat f$). Thus, $\hat a$ can have one of many arbitrary constant values instead of only one. The equations resulting from this action {\it do show} rudimentary, spin-like behavior in that the wave equations for $\psi $ and $\xi $ are equivalent to corresponding Dirac forms for constant, uniform, non-null electromagnetic fields when the symmetric part of the metric can be approximated as Lorentzian\cite{rankin.caqg}. Furthermore, an entire series of ``rest masses'' can now arbitrarily be introduced by choosing $\hat f$ to have a desired set of roots. Obviously, this is unsatisfactorily ad-hoc, but at least ``rest mass'' is not restricted to a single value. It has become dynamically determined by $\hat a$ as part of the solution. This result was actually used to investigate cosmological solutions, with an inflationary scenario for negative ``rest mass'' squared, and a ``Big Bang'' scenario for positive ``rest mass'' squared\cite{taylor.rankin}. The option to choose either case was essential for that dual behavior, although no realistic transition between the two cases was found. Indeed, only one example was found that seemed to allow any transition at all. The transition was assumed steplike, with continuity required in the stress tensor at the jump.

But yet another point can be made from this simple example. When the equations of motion are examined more closely, the constant $j^2 $ is found to always appear in the combination 
$j^2 \{1\mp [1/(2j^2 )][K\hat \beta \hat P^{-1/2}+CC]\} $, where 
$\hat P=\hat P_{\mu \nu }\hat P^{\mu \nu }$, 
$\hat P_{\mu \nu }=\hat p_{\mu \nu }+i\: {}^* \hat p_{\mu \nu }$, 
and $\hat f(K^2 )=0$. Thus for this case, the classical Poynting vector term\cite{jackson.self.force} is essentially replaced via
{\samepage 
\begin{eqnarray}
-j^2 (\vec E\times \vec B) & \to & -j^2 (\vec E\times \vec B)\pm 
{\textstyle{1 \over 4}}[(\hbar c)/e]\} \{ [K\hat \beta 
(\vec E\times \vec B)/(E^2-B^2 \nonumber \\
 &     & +2i\vec E{\bf \cdot}\vec B\, )^{1/2}\, ]+CC\} 
\label{replace.E.cross.B}
\end{eqnarray}}%

In regions containing electromagnetic sources, $\hat \beta \neq 0$, and the total electromagnetic field will be non-null\cite{jackson.self.force}, so the second term is well defined. But the primary point here is that in regions free of sources ($\hat \beta =0$), and containing null electromagnetic fields, the additional term is an indeterminate form like $0/0$. In a sense, this provides a classical, field theoretic example of uncertainty, which might at least display some of the paradoxical behavior associated with a photon in a two slit experiment. And in the vicinity of the final target, or the sources, this same term should be nonlinear in behavior.

This same combination of terms normally remains nonzero as 
$j \rightarrow 0$, which is the limit in which the standard, classical, Maxwell term in the action vanishes. The same sort of thing happens in the denominator of the four current in Maxwell's equations. Thus, a semblance of Maxwell's equations and the Maxwell stress tensor can survive in this example without the Maxwell term in the action, with its known disruptive tendencies\cite{jackson.self.force}. On the other hand, zeroes of 
$\hat \beta $ may then cause divergent values in the electromagnetic four current, and the vanishing of $j$ would completely destroy many of the parallels with Einstein-Maxwell-Schr\"{o}dinger fields in the earlier part of this paper. Thus, while the abandonment of the classical Maxwell term in the action at first seems esthetically pleasing, it is probably suicidal for the physics of the resulting theory. Such ``esthetic'' maneuvers are a constant temptation in classical unified field theories, but here at least, this one seems inadvisable. On the other hand, the entire example seems too arbitrary and ad-hoc, and is also unlikely to be very realistic. It has been used mainly to show the potential for unexpected new behavior in such theories. It also has produced a very limited agreement with standard Dirac theory\cite{rankin.caqg}. Any such clue could be worth some notice.

\subsection{Current Work - Torsion with a Tracefree Part}

If the antisymmetric part of the metric, $\hat a_{\mu \nu }$, is actually a key to extending the action of equation (\ref{complex.gi.action}), then it should have a firmer foundation. The model of equation (\ref{complex.gi.ahat.action}) simply imposes constancy on $\hat a$ with little more justification than convenience of results. On the other hand, in the much earlier works on asymmetric field theories, the constancy of the invariants of the antisymmetric part of the metric was actually an immediate result of some possible equations of motion. For example, Hlavat\'{y} notes in his study of Einstein's unified field theory that such a result will follow directly from {\it one possible} relation between metric and connection\cite{hlavaty}. This focuses particular attention on the relation between metric and connection.

But the connection $\Gamma ^\mu _{\nu \alpha }$ of a Riemann-Cartan geometry is defined simply by
\begin{equation}
g_{\mu \nu ,\alpha} -g_{\lambda \nu}\Gamma ^\lambda _{\mu \alpha }
-g_{\mu \lambda }\Gamma ^\lambda _{\nu \alpha }=0
\label{def.cartan.connect}
\end{equation}
where $g_{\mu \nu }$ is symmetric\cite{shtanov,rankin.caqg}. To jump from that to equation (\ref{def.connect.cartan}) requires an additional arbitrary constraint, namely that the tracefree part of the torsion vanishes
\cite{wheeler.decomp}. When the torsion $N^\mu _{\nu \alpha }$ is allowed a tracefree portion, 
$Q^\mu _{\nu \alpha }\equiv Q^{\; \; \; \; \: \mu }_{\nu \alpha }$ (to keep index positions unambiguous), equation (\ref{def.connect.cartan}) becomes
\begin{equation}
\Gamma ^\mu _{\nu \alpha }=\{ ^{\: \mu }_{\nu \alpha } \}+
\delta ^\mu _\alpha v_\nu -g_{\nu \alpha }v^\mu 
-Q^{\mu }_{\; \; \nu \alpha }-Q^{\mu }_{\; \; \alpha \nu }
+Q^{\; \; \; \; \: \mu }_{\nu \alpha }
\label{def.connect.cartan.q}
\end{equation}
Since the trace of the torsion is 
$N^\mu _{\nu \mu }={\textstyle{3 \over 2}}\, v_\nu $, the trace of the torsion absorbs the entire gauge freedom of the non-Christoffel part of the connection, and $Q^{\; \; \; \; \: \mu }_{\nu \alpha }=
\hat Q^{\; \; \; \; \: \mu }_{\nu \alpha }$\cite{rankin.caqg,shtanov}, or 
$Q^{\; \; \; \; \: \mu }_{\nu \alpha }$ is intrinsically gauge invariant.

Thus, this structure represents a simple extension of the earlier models, yet one that allows the addition of another field, $\hat Q^{\; \; \; \; \: \mu }_{\nu \alpha }$, which always had a logical right to be present. Furthermore, 
$\hat Q^{\; \; \; \; \: \mu }_{\nu \alpha }$ is to be determined through the action principle, so that equation (\ref{def.cartan.connect}) is supplemented by consistent additional equations to fully determine the connection. No terms involving $\hat Q^{\; \; \; \; \: \mu }_{\nu \alpha }$ contracted with itself or other quantities, and having dimensionless coupling constants, are introduced in the body of the action. Rather, 
$\hat Q^{\; \; \; \; \: \mu }_{\nu \alpha }$ is introduced solely through its role in the generalization of equation (\ref{gi.basic.identity}), which must be carried as a constraint in order to consistently vary the gauge invariant variables\cite{rankin.caqg}. Furthermore, the 
$j^2 \hat p_{\mu \nu }\hat p^{\mu \nu }$ term itself is not generalized to include $\hat a^{\mu \nu }$ terms, although such terms have an ``esthetic'' right to enter, and any contribution of $\hat a$ to the leading $\hat R$ term is assumed to cancel with the contribution of $\hat a$ to the generalization of the $\sqrt{-\hat g}$ term. The cosmological term is also not changed. In short, with one exception, no changes are made to equation (\ref{complex.gi.ahat.action}) that aren't absolutely required by 
$\hat Q^{\; \; \; \; \: \mu }_{\nu \alpha }\neq 0$ and the fact that it is tracefree, except to drop the constraint $\hat f(\hat a)=0$. The exception is the inclusion of one new dimensionless constant $k$ multiplying the leading $\hat R$ term for additional flexibility. Of course, these guidelines may require modification in the future.

Most of the details of this work will be reserved for a future presentation. However, the action just outlined is
{\samepage 
\begin{eqnarray}
I & = & {\textstyle{1 \over 2}}\int {\{ (k\hat R-2\sigma )-
{\textstyle{1 \over 2}}\, j^2(\hat p_{\mu \nu }\hat p^{\mu \nu })+
\hat \beta [\hat g^{\nu \alpha }\hat B_{\nu \alpha }-
\hat a^{\nu \alpha }\hat B_{\nu \alpha }}\nonumber \\
 &   & -e^{-i\hat \phi }(1+{\textstyle{1 \over 4}}\, \hat a)]
+2\hat \alpha ^{\mu \nu }\hat p_{\mu \nu }
+\hat \gamma ^\nu \hat Q^{\; \; \; \; \: \sigma }_{\nu \sigma }+CC\} 
\sqrt {-\hat g}\, d^4x
\label{gi.current.action}
\end{eqnarray}}%
where
{\samepage 
\begin{eqnarray}
\hat B_{\nu \alpha } & = & \hat R_{\nu \alpha }+2\hat g_{\nu \alpha }
\hat v^\sigma \hat v_\sigma +\hat v_{\alpha ||\nu }+\hat v_{\nu ||\alpha }
-\hat p_{\nu \alpha }-2\hat v_\nu \hat v_\alpha +\hat g_{\nu \alpha }
\hat v^\sigma _{\; \; ||\sigma }+
\hat Q^{\sigma }_{\; \; \nu \alpha ||\sigma }\nonumber \\
 &   & +\hat Q^{\sigma }_{\; \; \alpha \nu ||\sigma }
-\hat Q^{\; \; \; \; \: \sigma }_{\nu \alpha \; \; \, ||\sigma }
+2\hat v^\sigma \hat Q_{\sigma \nu \alpha }
-2\hat v^\sigma \hat Q_{\nu \alpha \sigma }
+2\hat Q^{\sigma \gamma}_{\; \; \; \; \nu }
\hat Q_{\gamma \alpha \sigma }
+\hat Q^{\sigma \gamma}_{\; \; \; \; \nu }
\hat Q_{\gamma \sigma \alpha }
\label{def.b.q}
\end{eqnarray}}%
and all the terms in the action of equation (\ref{gi.current.action}) are now placed inside the bracket containing the $CC$ term. This last change is not really required for some of the terms in this particular action. However, it could become necessary if $\hat a_{\mu \nu }$ is introduced into the terms which are real here.

When $\hat Q^{\; \; \; \; \: \mu }_{\nu \alpha }$ and the Lagrange multiplier $\hat \gamma ^\nu$ are varied as additional, complex quantities, keeping in mind that $\hat Q^{\; \; \; \; \: \mu }_{\nu \alpha }=
-\hat Q^{\; \; \; \; \: \mu }_{\alpha \nu }$, the resulting equations can be solved in general for $\hat Q^{\; \; \; \; \: \mu }_{\nu \alpha }$. The solution is
{\samepage 
\begin{eqnarray}
12\hat Q_{\mu \nu \alpha } & = & \hat a_{\nu \alpha ||\mu }-
\hat a_{\mu \alpha ||\nu }-2\hat a_{\mu \nu ||\alpha }
+\hat g_{\nu \alpha }\hat a^{\; \; \sigma }_{\mu \; \; ||\sigma }
-\hat g_{\mu \alpha }\hat a^{\; \; \sigma }_{\nu \; \; ||\sigma }
-\hat a_{\nu \alpha }\hat Z_\mu \nonumber \\
 &   & +\hat a_{\mu \alpha }\hat Z_\nu +2\hat a_{\mu \nu }\hat Z_\alpha 
-\hat g_{\nu \alpha }\hat a_{\mu \sigma }\hat Z^\sigma 
+\hat g_{\mu \alpha }\hat a_{\nu \sigma }\hat Z^\sigma 
-12\sqrt {-\hat g}\, \epsilon_{\mu \nu \alpha \lambda }\hat W^\lambda 
\label{q.general.solution}
\end{eqnarray}}%
where
\begin{equation}
\hat Z_\mu =3\hat v_\mu -i3\hat W_\mu -(ln\, \hat \beta )_{,\mu }
\label{def.z}
\end{equation}
and
{\samepage 
\begin{eqnarray}
3(8+\hat a)\hat W^\lambda & = & -i4\hat a^{\lambda \sigma }
(ln\, \hat \beta )_{,\sigma }+i\hat a\hat g^{\lambda \sigma }
(ln\, \hat \beta )_{,\sigma }+i{\textstyle{3 \over 4}}\, 
\hat g^{\lambda \sigma }\hat a_{,\sigma }\nonumber \\
 &   & -i3\hat a\hat v^\lambda 
-i4\hat a^{\lambda \sigma }_{\; \; \; \; ||\sigma }+
i2\hat a^\lambda _{\; \; \tau }\hat a^{\tau \sigma }_{\; \; \; \; ||\sigma }
\label{def.w}
\end{eqnarray}}%
Here, $\hat W^\lambda $ is actually the dual of the totally antisymmetric part of $\hat Q_{\mu \nu \alpha }$, so it is technically a pseudovector. However, the presence of the self dual tensor $\hat a_{\mu \nu }$ gives peculiar properties to $\hat Q_{\mu \nu \alpha }$ itself, so even this point is complicated somewhat by the fact that $\hat a_{\mu \nu }$ contains a dual internally\cite{rankin.caqg}. Note that it is assumed in proposing any model like this, that the quantity $(ln\, \hat \beta )_{,\sigma }$ will not disrupt observables. A similar requirement was necessary (but unstated) for 
$(ln\, \psi )_{,\mu }$ in $\hat v_\mu $ in the first part of this paper, although it was obviously satisfied there. In both cases, the immediate quantity may diverge at points. However, $\hat v_\mu $ caused no divergences in the observables in the earlier model because of accompanying factors containing $\hat \beta $, or even simply $\psi $. Here, the worst that seems possible is that the zeroes of $\xi $ and $\psi $ would be required to coincide to ensure well behaved results. Such restrictions would not seem extreme from the viewpoint of standard quantum theory.

There are also cases where $\hat Q_{\nu \alpha \mu }$ is decomposed into a totally antisymmetric part, and a residual portion, and either of these parts is constrained to vanish. If the residual portion is constrained to vanish so that the tracefree part of the torsion is totally antisymmetric, equation (\ref{q.general.solution}) is replaced by
\begin{equation}
\hat Q_{\mu \nu \alpha }=-\sqrt {-\hat g}\, 
\epsilon_{\mu \nu \alpha \lambda }\hat W^\lambda 
\label{q.antisymmetric}
\end{equation}
where
\begin{equation}
6\hat W^\lambda =-i\hat a^{\lambda \sigma }_{\; \; \; \; ||\sigma }-
i\hat a^{\lambda \sigma }(ln\, \hat \beta )_{,\sigma }
\label{def.w.antisymmetric}
\end{equation}
This constrained form of the model can be contrasted with some of the results of a theory of torsion and spin developed by Hammond\cite{hammond}.

While few implications of the above forms are yet known, notice that 
$\hat Q^{\; \; \; \; \: \mu }_{\nu \alpha }=0$ if 
$\hat a_{\mu \nu }=0$. This not only provides a belated justification for excluding a tracefree part to the torsion in the Weyl-Cartan case of equation (\ref{complex.gi.action}), but it also gives some indirect support to the idea that this approach may be on the right track once it gets the geometric model fully consistent internally. The ``spin zero'' Weyl-Cartan model correctly reproduces the relativistic Schr\"{o}dinger equation form, and it is also the model in which even the relation between metric and connection is consistent with the simple action principle given, plus the requirement of metric compatibility. Hopefully, the extended model has simply had the details of the kinematics and the action wrong once one leaves ``spin zero'' behind. It does appear true however, that even ``spin zero'' requires some of these enhancements to handle more than one rest mass.

There are also some reasons to believe that a tracefree part to the torsion may relate to SU(2) type of interactions. If a spinor connection corresponding to the connection of equation (\ref{def.connect.cartan.q}) (or actually, its gauge invariant analog) is constructed\cite{carmeli}, the result may be reorganized so that it contains a set of terms that have the form of SU(2) potentials constructed from elements of 
$\hat Q^{\; \; \; \; \: \mu }_{\nu \alpha }$ and its dual taken on its first two indices\cite{rankin.mg7}. However, the fact that equation (\ref{def.connect.cartan.q}) defines a presumably complex connection, raises some fine points in the  translation to spin space.

As negatives, this newer extended formalism has yet to demonstrate even the limited agreements with Dirac theory produced by the earlier model with an antisymmetric part to the metric\cite{rankin.caqg}. Furthermore, there is yet no evidence that the quantity $\hat a$ behaves in a manner suitable for incorporation into the rest mass term once it is unconstrained. Indeed, it should be able to pivot freely in the complex plane, and thus act somewhat like a complex scattering potential in quantum theory\cite{schiff}, but with no loss of conservation laws. One can then arbitrarily forbid ``pathological'' behavior as is usually done with quantum mechanical wave equations, but it is not yet known if this will lead to any useful conditions on $\hat a$ or the ``rest mass'' parameter. However, imposition of arbitrary constraints on $\hat a$ in the action is not very satisfactory either, since the ``rest masses'' then become preset artificially by hand, and there is still no guarantee that more general agreement with Dirac theory will appear. Thus, this newer extended model remains quite speculative and tentative, although there may be some merit to the overall idea of adding in an antisymmetric part to the metric related to spin, and a tracefree part to the torsion to obtain necessary, additional interactions.

\section{Conclusions}
\label{conclusions}

This paper has demonstrated the equivalence between using certain, simple gauge invariant variables, and using more standard, gauge dependent, Weyl type variables in action principles for Weyl-like geometries with 
nonzero scalar curvature. Furthermore, the gauge invariant variables have been shown to provide several advantages over their gauge dependent counterparts, both for formulating action principles, and for interpreting the results. This demonstration has been a primary goal of this paper, since the gauge invariant variables automatically obey an identity which is often mistaken for a gauge condition. That particular identity has been given careful treatment in this paper to dispose of that misconception. Additionally, the complete gauge freedom of the equations has been underscored by providing the solution they give for the Weyl metric via equation (\ref{weyl.metric.solution}), a solution which still explicitly contains the appropriate gauge freedom for that metric.

As a second point, use of gauge invariant variables highlights the intrinsic Schr\"{o}dinger-like tendencies of Weyl-like geometries. Such behavior has attracted attention since the early days of quantum mechanics\cite{london.orig,adler.london}, and I feel that it provides a primary reason for interest in Weyl-like theories, including the use of gauge invariant variables to examine them. In particular, the choice of an Einstein-Maxwell action with an imaginary Maxwell field, leads automatically to the appearance of a pair of relativistic Schr\"{o}dinger fields in all the field equations. Therefore the possible relation between such behavior in this model, and the roots of quantum phenomena has been discussed, along with difficulties with this approach, including possible negative energy densities, and the need to demonstrate particle behavior. Some new features which may be relevant to these problems have also been mentioned, particularly the nonlinearity of the wave equations for strong fields, and the feedback effect of the scalar curvature on the stress tensor magnitude in strong fields. The introduction into the geometry of a self dual, antisymmetric part to the metric and a tracefree part to the torsion, has been outlined, and a particular action under current investigation has been summarized. One of the possible goals of such an approach is an understanding of spin. An understanding of the ``rest mass'' parameter in the wave equations may also be linked to this, and these newer, geometric components provide some room to introduce additional interactions into the study. However, these extended formalisms have been noted to be less ``natural'' than the simpler model of the earlier part of the paper, at least so far.

\appendix* 
\section{Coupled Einstein-Maxwell-Klein-Gordon Fields}
\label{norm.kg}

The conventional theory of a coupled, Einstein-Maxwell-Klein-Gordon field can be derived in the standard manner from the action\cite{morse.feshbach,%
wald,hawking.ellis}
{\samepage 
\begin{eqnarray}
I & = & \frac{1}{2kc}\int {\{(R-2\Lambda )-[k/(8\pi )]F_{\mu \nu }
F^{\mu \nu }+[(k\hbar c)/m][g^{\mu \nu }\, {}^\# (\psi _{,\mu }}
\nonumber \\
  &   & +iqA_\mu \psi )(\psi _{,\nu }+iqA_\nu \psi )-m^2 \, {}^\# \psi 
\psi ]\}\sqrt {-g}\, d^4x
\label{stan.kg.action}
\end{eqnarray}}%
where $k=8\pi (G/c^4 )$, $q=e/(\hbar c)$, and $m=(m_0 c)/\hbar $. When treated as a functional of $g^{\mu \nu }$,$A_\mu $, $\psi $, and 
${}^\# \psi $, this action principle gives
\begin{equation}
F^{\mu \nu }_{\; \; \; \; ;\nu }=
2\pi [(iq\hbar c)/m]g^{\mu \nu }\{\, {}^\# \psi _{,\nu }\psi - 
\psi _{,\nu }\, {}^\# \psi -2iq\, {}^\# \psi \psi A_\nu \}
\label{reg.sch.maxwell.result}
\end{equation}
{\samepage 
\begin{eqnarray}
R_{\mu \nu }-{\textstyle{1 \over 2}}\, Rg_{\mu \nu }+
\Lambda g_{\mu \nu } & = & -\frac{k}{4\pi }(F_\mu ^{\; \; \gamma }
F_{\gamma \nu }+{\textstyle{1 \over 4}}\, g_{\mu \nu }
F_{\gamma \tau }F^{\gamma \tau })\nonumber \\
 &   & -\frac{k\hbar c}{2m}[\, {}^\# (\psi _{,\mu }+iq\psi A_\mu )
(\psi _{,\nu }+iq\psi A_\nu )\nonumber \\
 &   & +\, {}^\# (\psi _{,\nu }+iq\psi A_\nu )(\psi _{,\mu }+
iq\psi A_\mu )-\, {}^\# (\psi _{,\gamma }\nonumber \\
 &   & +iq\psi A_\gamma )(\psi _{,\tau }+iq\psi A_\tau )
g^{\gamma \tau }g_{\mu \nu }+m^2 \, {}^\# \psi \psi g_{\mu \nu }] 
\label{reg.sch.einstein.result}
\end{eqnarray}}%
and
{\samepage 
\begin{eqnarray}
& (1/\sqrt {-g}\, ) & (\sqrt {-g}\, g^{\mu \nu }
\psi _{,\nu })_{,\mu }+2iqg^{\mu \nu }A_\mu \psi _{,\nu }\nonumber \\
&  & +[(iq)/\sqrt {-g}\, ](\sqrt {-}\, g^{\mu \nu }A_\nu )_{,\mu }\psi 
-q^2 g^{\mu \nu }A_\mu A_\nu \psi =-m^2 \psi 
\label{reg.wave.eq}
\end{eqnarray}}%
These standard equations should be compared to equations (\ref{main.wave.eq}), (\ref{imag.sch.maxwell.result.2}), and (\ref{imag.sch.einstein.result.2}) of this current paper.

Note that the Klein-Gordon field energy density is positive definite for this action above, as is the Maxwell energy density. As noted in section \ref{quantum.discussion}, the formalism of this paper actually resolves into a pair of Klein-Gordon fields, one of which has negative energy density. A ``standard'' form action based on the above, and which achieves this same type of result, would be
{\samepage 
\begin{eqnarray}
I & = & \frac{1}{2kc}\int {\{(R-2\Lambda )-[k/(8\pi )]F_{\mu \nu }
F^{\mu \nu }+\sum _{n=1}^2 (-1)^n [(k\hbar c)/m][g^{\mu \nu }\, {}^\# 
(\zeta _{n,\mu }}\nonumber \\
  &   & +iqA_\mu \zeta _n )(\zeta _{n,\nu }+iqA_\nu \zeta _n )-m^2 
\, {}^\# \zeta _n \zeta _n ]\}\sqrt {-g}\, d^4x
\label{dual.kg.action}
\end{eqnarray}}%
Here, both $\zeta _1 $ and $\zeta _2 $ are Klein-Gordon fields just as 
$\psi $ was in equation (\ref{stan.kg.action}).

Finally, Wald discusses allowing a non-minimal interaction between the 
Klein-Gordon field and gravitation of the form $(1/6) R\psi $ in equation (\ref{reg.wave.eq})\cite{wald}. Such a term would allow the action of equation (\ref{dual.kg.action}) to generalize still further to
{\samepage 
\begin{eqnarray}
I & = & \frac{1}{2kc}\int {\{(R-2\Lambda )-[k/(8\pi )]F_{\mu \nu }
F^{\mu \nu }+\sum _{n=1}^2 (-1)^n [(k\hbar c)/m][g^{\mu \nu }\, {}^\# 
(\zeta _{n,\mu }}\nonumber \\
  &   & +iqA_\mu \zeta _n )(\zeta _{n,\nu }+iqA_\nu \zeta _n )-(m^2 
+{\textstyle{1 \over 6}}\, R)\, {}^\# \zeta _n \zeta _n ]\}
\sqrt {-g}\, d^4x
\label{dual.conf.kg.action}
\end{eqnarray}}%
Except for factors which can be absorbed by changing to dimensionless coordinates via the scale factor $b_0 ^{-1/2}$, the recognition that the metric is (already) the gauge invariant metric, and the denormalization of the $\zeta _n $ to absorb the factor containing $G$, this action should give equations of motion which are the same as equations (\ref{zeta.sch.maxwell.result}), (\ref{zeta.sch.einstein.result}), (\ref{zeta.rhat}), and (\ref{zeta.nonlin.wave.eq}).

\begin{acknowledgments}

I would like to thank Jim Wheeler, Daniel Galehouse, Bill Baker, Egon Marx, and David Finkelstein for discussions and points that have contributed significantly to this paper. I would also like to thank Frank Taylor, who coauthored some earlier conference contributions on cosmological applications of this approach with me.

\end{acknowledgments}



\begin{thebibliography}{99}
\bibitem{wheeler.scalars}  Wheeler makes the point that only gauge invariant scalars, or scalars of gauge weight zero, can truly correspond to observables\cite{wheeler.physrev}.
\bibitem{galehouse.minus.C}  My thanks to Daniel Galehouse for correctly insisting that cases which correspond to negative values for C are legitimate.
\end{thebibliography}
\end{document}